\pdfoutput=1

\newif\ifFull
\newif\ifShort
\newif\ifArxiv
\Fulltrue
\ifFull
\Arxivtrue 
\Shortfalse
\documentclass[pdftex,11pt]{article}
\advance \topmargin by -\headheight
\advance \topmargin by -\headsep
\evensidemargin \oddsidemargin
\usepackage{amsmath}
\usepackage{amssymb}
\else
\documentclass[pdftex]{vldb}
\Shorttrue
\fi
\usepackage{graphicx}
\usepackage{url}
\usepackage{subfigure}
\usepackage{hyperref}

\ifShort
\usepackage{balance}
\fi   

\newcommand{\Comment}[1]{\relax}

\newcommand{\I}{\mathcal{I}}
\newcommand{\T}{\mathcal{T}}
\newcommand{\W}{\mathcal{W}}
\newcommand{\F}{\mathcal{F}}

\begin{document}

\ifShort
\title{Efficient Verification of Web-Content Searching \\Through
Authenticated Web Crawlers%
}
\else
 \title{Verifying Search Results Over Web Collections}
\fi

\ifFull
\author{%
  \makebox[.3\linewidth]{Michael T.~Goodrich}\\ UC Irvine\\
  \and \makebox[.3\linewidth]{Duy Nguyen}\\ Brown University\\
  \and \makebox[.2\linewidth]{Olga Ohrimenko}\\Brown University\\
  \and \makebox[.3\linewidth]{Charalampos Papamanthou}\\ UC Berkeley\\
  \and \makebox[.3\linewidth]{Roberto Tamassia}\\Brown University\\
  \and \makebox[.5\linewidth]{Nikos Triandopoulos}\\RSA Laboratories \& Boston University\\
  \and \makebox[.5\linewidth]{Cristina Videira Lopes}\\UC Irvine\\
}
\date{}
\fi

\maketitle
\ifShort
\input{short_abstract}
\else
\begin{abstract}
  Web searching accounts for one of the most frequently performed
  computations over the Internet as well as one of the most important
  applications of outsourced computing, producing results that
  critically affect users' decision-making behaviors. As such, verifying
  the integrity of Internet-based searches over vast amounts of web
  contents is essential.
 
  In this paper, we provide the first solution to this general security
  problem.  We introduce the concept of an \emph{authenticated web
    crawler} and present the design and prototype implementation of this
  new concept. An authenticated web crawler is a trusted program that
  computes a special ``signature'' $\mathcal{S}$ over a collection of web
  contents it visits. Subject to this signature, web searches can be
  verified to be correct with respect to the integrity of their produced
  results.  But this signature serves more advanced purposes than just
  \emph{content verification}: It allows the verification of complicated
  queries on web pages, such as \emph{conjunctive keyword searches},
  which are vital for the functionality of online web-search engines
  today. In our solution, along with the web pages that satisfy any
  given search query, the search engine also returns a cryptographic
  proof. This proof, together with the signature $\mathcal{S}$, enables
  any user to efficiently verify that no legitimate web pages are
  omitted from the result computed by the search engine, and that no
  pages that are non-conforming with the query are included in the
  result. An important property of our solution is that the proof size
  and the verification time are proportional only to the sizes of the
  query description and the query result, but do not depend on the
  number or sizes of the web pages over which the search is performed.  

  Our authentication protocols are based on standard Merkle trees and
  the more involved bilinear-map accumulators.  As we experimentally
  demonstrate, the prototype implementation of our system gives a low
  communication overhead between the search engine and the user, and
  allows for fast verification of the returned results on the user side.
\end{abstract}


\fi

\ifFull
\thispagestyle{empty}
\fi

\section{Introduction}\label{sec:intro}

\ifFull
It goes without saying that web searching is an essential part of modern
life.
\fi
When we perform a web search, we expect that the list of links returned
will be relevant and complete.
As we heavily rely on web searching, an often overlooked issue is that
search engines are outsourced computations.  That is, users issue
queries and have no intrinsic way of trusting the results they receive,
thus introducing a modern spin on Cartesian doubt.  This philosophy once
asked if we can trust our senses---now it should ask if we can trust our
search results.  Some possible attack scenarios that arise in this
context include the following:
\begin{enumerate}
\item A news web site posts a misleading article and later changes it to
  look as if the error never occurred.
\item A company posts a back-dated white paper claiming an invention
  after a related patent is issued to a competitor.
\item An obscure scientific web site posts incriminating data about a
  polluter, who then sues to get the data removed, in spite of its
  accuracy.
\item A search engine censors content for queries coming from users in a
  certain country, even though an associated web crawler provided web
  pages that would otherwise be indexed for the forbidden queries.
\end{enumerate}
An Internet archive, such as in the Wayback Machine, that digitally
signs the archived web pages could be a solution to detecting the first
attack, but it does not address the rest.  It misses detecting the
second, for instance, since there is no easy way in such a system to
prove that something did not exist in the past. Likewise, it does not
address the third, since Internet archives tend to be limited to popular
web sites.  Finally, it does not address the fourth, because such users
would likely also be blocked from the archive web site and, even
otherwise, would have no good way of detecting that pages missing from a
keyword-search response.

From a security point of view, we can abstract these problems in a model
where a query request (e.g., web-search terms) coming from an end user,
Alice, is served by a remote, unknown and possibly untrusted server
(e.g., online search engine), Bob, who returns a result consumed by
Alice (e.g., list of related web pages containing the query terms).  In
this context, it is important that such computational results are
\emph{verifiable by the user}, Alice, with respect to their
\emph{integrity}.  Integrity verifiability, here, means that Alice
receives additional authentication information (e.g., a digital
signature from someone she trusts) that allows her to verify the
integrity of the returned result. In addition to \emph{file-level
  protection}, ensuring that data items (e.g., web contents) remain
intact, the integrity of the returned results typically refers to the
following three properties (e.g.,~\cite{lhkr-dai-06}): (1)
\emph{correctness}, ensuring that any returned result satisfies the
query specification; (2) \emph{completeness}, ensuring that no result
satisfying the query specification is omitted from the result, and (3)
\emph{freshness}, ensuring that the returned result is computed on the
currently valid, and most recently updated data.

The ranking of query results is generally an important part of web
searching. However, in the above scenarios, correctness, completeness,
and freshness are more important to the user than a proof that the
ranking of the results is accurate. Also, note that it is usually in the
best interest of the search engine to rank pages correctly, e.g., for
advertising.
Thus in our context, we are interested in studying the problem of
\emph{integrity verification} of web content.  In particular, we wish to
design protocols and data-management mechanisms that can provide the
user with a cryptographically verifiable proof that web content of
interest or query results on this content are \emph{authentic},
satisfying \emph{all} the above three security properties: correctness,
completeness, and freshness.

\subsection{Challenges in Verifying Web Searching}
Over the last decade, significant progress has been made on integrity
protection for \emph{management} \emph{of} \emph{outsourced}
\ifShort
\emph{data-bases}.
\else
\emph{databases}.
\fi
Here, a database that is owned by a (trusted) source, Charles, is
outsourced to an (untrusted) server, Bob, who serves queries coming from
end users such as Alice.  Using an \emph{authenticated index structure},
Bob adds \emph{authentication} to the supported database responses, that
is, augments query results with authentication information, or
\emph{proof}, such that these results can be cryptographically verified
by Alice with respect to their integrity.  This authentication is done
subject to a \emph{digest} that the source, Charles, has produced by
processing the currently valid database version. In the literature, many
elegant solutions have been proposed focusing on the authentication of
large classes of queries via short proofs (sent to the user) yielding
fast verification times.

Unfortunately, translating such existing methods in the authenticated
web searching problem is not straightforward, but rather entails several
challenges. First, result verifiability is not well defined for web
searching, because unlike the database outsourcing model, in web
searching there is no clear data source and there is no real data
outsourcing by someone like Charles.  Of course, one could consider a
model where each web-page owner communicates with the online search
engine to deliver (current) authentication information about the web
pages this owner controls, but clearly this consideration would be
unrealistic.
Therefore, we need an authentication scheme that is \emph{consistent
  with the crawling-based current practices of web searching}.

Additionally, verifying the results of search engines seems particularly
challenging given the large scale of this data-processing problem and
the high rates at which data evolves over time. Indeed, even when we
consider the static version of the problem where some portion of the web
is used for archival purposes only (e.g., older online articles of the
Wall Street Journal), authenticating general search queries for such
rich sets of data seems almost impossible: How, for instance, is it
possible to verify the completeness of a simple keyword-search
query over a large collection of archival web pages?  Existing
authentication methods heavily rely on a total ordering of the database
records on some (primary-key) attribute in order to provide
``near-miss'' proofs about the completeness of the returned records, but
no such total order exists when considering keyword-search queries over
text documents. This suggests that to prove completeness to a user the
search engine will have to provide the user with ``all supporting
evidence''---all web contents the engine searched through---and let the
user recompute and thus trivially verify the returned result. Clearly,
this approach is also not scalable.  Generally, Internet searching is a
complicated ``big data'' problem and so is its authentication: We thus
need an authentication scheme that produces proofs and incurs
verification times that are \emph{web-search sensitive}, that is, they
\emph{depend on the set of returned results and not on the entire
  universe of possible documents}.

Integrity protection becomes even more complicated when we consider web
pages that frequently change over time.  For instance, collaborative
systems store large amounts of scientific data at distributed web-based
repositories for the purpose of sharing knowledge and data-analysis
results. Similarly, users periodically update their personal or
professional web pages and blogs with information that can be searched
through web search engines. In such dynamic settings, how is it possible
to formally define the notion of freshness? Overall, we need an
authentication scheme that is \emph{consistent with the highly dynamic
  nature of web content}.

\subsection{Prior Related Work}
Despite its importance and unlike the well-studied problem of database
authentication, the problem of web searching authentication has not been
studied before in its entirety. To the best of our knowledge, the only
existing prior work studies a more restricted version of the problem.

The first solution on the authentication of web searches was recently
proposed by Pang and Mouratidis in PVLDB 2008~\cite{pm-aqrtse-08}.  This
work focuses on the specific, but very important and representative
case, where search engines perform \emph{similarity-based document
  retrieval}. Pang and Mouratidis show how to construct an
authentication index structure that can be used by an untrusted server,
the search engine, in the \emph{outsourced database model} to
authenticate text-search queries over documents that are based on
similarity searching. In their model, a trusted owner of a collection of
documents outsources this collection to the search engine, along with
this authentication index structure defined over the document
collection. Then, whenever a user issues a text-search query to the
engine, by specifying a set of keywords, the engine returns the top $r$
results ($r$ being a system parameter) according to some well-defined
notion of relevance that relates query keywords to documents in the
collection. In particular, given a specific term (keyword) an inverted
list is formed with documents that contain the term; in this list the
documents are ordered according to their estimated relevance to the
given term.
Using the authentication index structure, the engine is able to return
those $r$ documents in the collection that are better related to the
terms that appear in the query, along with a proof that allows the user
to verify the correctness of this document retrieval.


At a technical level, Pang and Mouratidis make use of an elegant
hash-based authentication method: The main idea is to apply
cryptographic hashing in the form of a Merkle hash tree
(MHT)~\cite{m-cds-89} over each of the term-defined lists. Note that
each such list imposes a total ordering over the documents it contains
according to their assigned score, therefore completeness proofs are
possible. Pang and Mouratidis observe that the engine only partially
parses the lists related to the query terms: At some position of a list,
the corresponding document has low enough cost that does not allow
inclusion in the top $r$ results. Therefore, it suffices to provide
hash-based proofs (i.e., consisting of a set of hash values, similar to
the proof in a Merkle tree) only for \emph{prefixes} of the lists
related to the query terms. Pang and Mouratidis thus construct a novel
\emph{chained sequence of Merkle trees}. This chain is used to
authenticate the documents in an inverted list corresponding to a term,
introducing a better trade-off between verification time and size of
provided proof.

To answer a query, the following process is used.  For each term,
documents are retrieved sequentially through its document list, like a
sliding window, and the scores of these documents are aggregated. A
document's score is determined by the frequency of each query term in
the document. This parsing stops when it is certain that no document
will have a score higher than the current aggregated score.

To authenticate a query, the engine collects, as part of the answer and
its associated proof, a verification object that contains the $r$
top-ranked documents and their corresponding MHT proofs, as well as all
the documents in between that did not score higher but were potential
candidates. For example, consider the following two lists:
\begin{eqnarray*}
  &\mathsf{term}_1:&\mathsf{doc}_1, \mathsf{doc}_5, \mathsf{doc}_3, \mathsf{doc}_2\\ 
  &\mathsf{term}_2:&\mathsf{doc}_2, \mathsf{doc}_5, \mathsf{doc}_3, \mathsf{doc}_4, \mathsf{doc}_1
\end{eqnarray*}
If a query is ``$\mathsf{term}_1~\mathsf{term}_2$'', $r$ is $1$ and
$\mathsf{doc}_1$ has the highest score, then the verification object has
to contain $\mathsf{doc}_2$, $\mathsf{doc}_5$, $\mathsf{doc}_3$,
$\mathsf{doc}_4$ to prove that their score is lower than
$\mathsf{doc}_1$.

We thus observe four limitations in the work by Pang and Mouratidis: (1)
their scheme is not fully consistent with the crawling-based web
searching, since it operates in the outsourced database model; (2) their
scheme is not web-search sensitive because, as we demonstrated above, it
returns documents and associated proofs that are not related to the
actual query answer, and these additional returned items may be of size
proportional to the \emph{number of documents in the collection}; (3)
their scheme requires complete reconstruction of the authentication
index structure when the document collection is updated: even a simple
document update may introduce changes in the underlying document scores,
which in the worst case will completely destroy the ordering within one
or more inverted lists; (4) it is not clear whether and how their scheme
can support query types different from disjunctive keyword searches.

\subsection{Our Approach}
Inspired by the work by Pang and Mouratidis~\cite{pm-aqrtse-08}, we
propose a new model for performing keyword-search queries over online
web contents.  In Section~\ref{sec:model}, we introduce the concept of
an \emph{authenticated web crawler}, an authentication module that
achieves authentication of general queries for web searching in a way
that is consistent with the current crawling-based search
engines. Indeed, this new concept is a program that like any other web
crawler visits web pages according to their link structure. But while
the program visits the web pages, it incrementally builds a
space-efficient authenticated data structure that is specially designed
to support general keyword searches over the contents of the web
pages. Also, the authenticated web crawler serves as a trusted component
that computes a signature, an accurate security snapshot, of the current
contents of the web. When the crawling is complete, the crawler
publishes the signature and gives the authenticated data structure to
the search engine. The authenticated data structure is in turn used by
the engine to provide verification proofs for any keyword-search query
that is issued by a user, which can be verified by having access to the
succinct, already published signature.

In Section~\ref{sec:solution}, we present our authentication methodology
which provides proofs that are web-search sensitive, i.e., of size that
depends linearly on the query parameters and the results returned by the
engine and logarithmically on the size of the inverted index (number of
indexed terms), but it does not depend on the size of the document
collection (number of documents in the inverted lists). The verification
time spent by the user is also web-search sensitive: It depends only on
the query (linearly), answer (linearly) and the size of the inverted index (logarithmically), not on the size of the web
contents that are accessed by the crawler. We stress the fact that our
methodology can support general keyword searches, in particular,
\emph{conjunctive keyword searches}, which are vital for the
functionality of online web search engines today and, accordingly, a
main focus in our work.

Additionally, our authentication solution allows for efficient updates
of the authenticated data structure. That is, if the web content is
updated, the changes that must be performed in the authenticated data
structure are only specific to the web content that changes and there is
no need to recompute the entire authentication structure. This property
comes in handy in more than one flavors: Either the web crawler itself
may incrementally update a previously computed authentication structure
the next time it crawls the web, or the crawler may perform such an
update on demand without performing a web crawling.

So far we have explained a three-party model where the client verifies
that the results returned by the search engine are consistent with the
content found by the web crawler. (See also
Section~\ref{sec:three-party-model} for more details). Our solution can
also be used in other common scenarios involving two parties. First,
consider a client that outsources its data to the cloud (e.g., to Google
Docs or Amazon Web Services) and uses the cloud service to perform
keyword-search queries on this data.  Using our framework, the client
can verify that the search results were correctly performed on the
outsourced data.  In another scenario, we have a client doing keyword
searches on a trusted search engine that delivers the search results via
a content delivery network (CDN).  Here, our solution can be used by the
client to discover if the results from the search engine were tampered
with during their delivery, e.g., because of a compromised server in
the~CDN.  (See Section~\ref{sec:other_models} for more details on these
two-party models).

Our authentication protocols are based on standard Merkle trees as well
as on bilinear-map accumulators. The latter are cryptographic primitives
that can be viewed as being equivalent to the former, i.e., they provide
efficient proofs of set membership. But they achieve a different
trade-off between verification time and update time: Verification time
is constant (not logarithmic), at the cost that update time is no more
logarithmic (but still sublinear). However, bilinear-map accumulators
offer a unique property not offered by Merkle trees: They can provide
constant-size proofs (and corresponding verification times) for
\emph{set disjointness}, i.e., in order to prove that two sets are
disjoint a constant-size proof can be used---this property is very
important for proving completeness over the unordered documents that are
stored in an inverted index
%
(used to support keyword searches).

In Section~\ref{sec:performance}, we describe implementation details of
the prototype of our solution and present empirical evaluation of its
performance on the Wall Street Journal archive.  We demonstrate that in
practice our solution gives a low communication overhead between the
search engine and the user, and allows for fast verification of the
returned results on the user side.  We also show that our prototype can
efficiently support updates to the collection.
We conclude in Section~\ref{sec:conclusion}.

\subsection{Additional Related Work}\label{sec:related}
\ifFull 
Before we describe the details of our approach, we briefly discuss some
additional related work.
\fi

A large body of work exists on \emph{authenticated data structures}
(e.g.,~\cite{dgms-adpoi-03,nn-crcu-98,ptt-ovods-11}), which provide a
framework for designing practical methods for query authentication:
Answers to queries on a data structure can be verified efficiently
through the computation of some short cryptographic proofs.

Research initially focused on authenticating membership
queries~\cite{bll-acmua-00} and the design of various authenticated
dictionaries~\cite{GoodrichTS01,nn-crcu-98,tt-cbhdp-05} based on
extensions of Merkle's \emph{hash tree}~\cite{m-cds-89}.  Subsequently,
one-way accumulators~\cite{cam-lys-02,nguyen05accumulators} were
employed to design dynamic authenticated
dictionaries~\cite{gth-aeddca-02,ptt-aht-08} that are based on algebraic
cryptographic properties to provide optimal verification overheads.

More general queries have been studied as well. Extensions of hash trees
have been used to authenticate various types of queries, including basic
operations (e.g., select, join) on databases~\cite{dgms-adpoi-03},
pattern matching in tries~\cite{mngdks-gmads-04} and orthogonal range
searching~\cite{ack-grid-08,mngdks-gmads-04}, path queries and
connectivity queries on graphs and queries on geometric
objects~\cite{gtt-eadsgcgsp-11} and queries on XML
documents~\cite{bcftg-satdxd-04a,dgkmns-faxd-01}.  Many of these queries
can be reduced to one-dimensional range-search queries which can been
verified optimally in~\cite{gtt-sevdod-08,n-hat-05} by combining
collision-resistant hashing and one-way accumulators.  Recently, more
involved cryptographic primitives have been used for optimally verifying
set operations~\cite{ptt-ovods-11}.

Substantial progress has also been made on the design of generic
authentication techniques. In~\cite{mngdks-gmads-04} it is shown how to
authenticate in the static case a rich class of search queries in DAGs
(e.g., orthogonal range searching) by hashing over the search structure
of the underlying data structure.  In~\cite{gtt-eadsgcgsp-11}, it is shown
how extensions of hash trees can be used to authenticate decomposable
properties of data organized as paths (e.g., aggregation queries over
sequences of objects) or any search queries that involve iterative
searches over catalogs (e.g., point location). Both works involve proof
sizes and verification times that asymptotically equal the complexity of
answering queries. Recently, in~\cite{tt-tda-10}, a new framework is
introduced for authenticating general query types over structured data
organized and accessed in the relational database model. By decoupling
the processes of query answering and answer verification, it is shown
how any query can be reduced (without loss of efficiency) to the
fundamental problem of set-membership authentication, and that
super-efficient answer verification (where the verification of an answer is
asymptotically faster than the answer computation) for many interesting
problems is possible. Set-membership authentication via
collision-resistant hashing is studied in~\cite{tt-cbhdp-05} where it is
shown that for hash-based authenticated dictionaries of size $n$, all
costs related to authentication are at least logarithmic in $n$ in the
worst case.

Finally, a growing body of works study the specific problem of
authenticating SQL queries over outsourced relational databases
typically in external-memory data management settings. Representatives
of such works include the authentication of range search queries by
using hash trees (e.g.,~\cite{dgms-adpoi-03,gtt-eadsgcgsp-11}) and by
combining hash trees with accumulators
(e.g.,~\cite{gtt-sevdod-08,n-hat-05}) or B-trees with signatures
(e.g.,~\cite{nt-aod-06,pjrt-vc-05}). Additional work includes an
efficient hash-based B-tree-based authenticated indexing technique
in~\cite{lhkr-dai-06}, the authentication of join queries
in~\cite{auth-join09} and of shortest path queries~in~\cite{mouratidis-shortest-paths}.



\section{Our model}
\label{sec:model}

We first describe two models where our results can be applied allowing
users to be able to verify searches over collections of documents of
interest (e.g., a set of web pages).

\subsection{Three-party Model}\label{sec:three-party-model}

\ifFull
\begin{figure}
\centering
\includegraphics[scale=0.5]{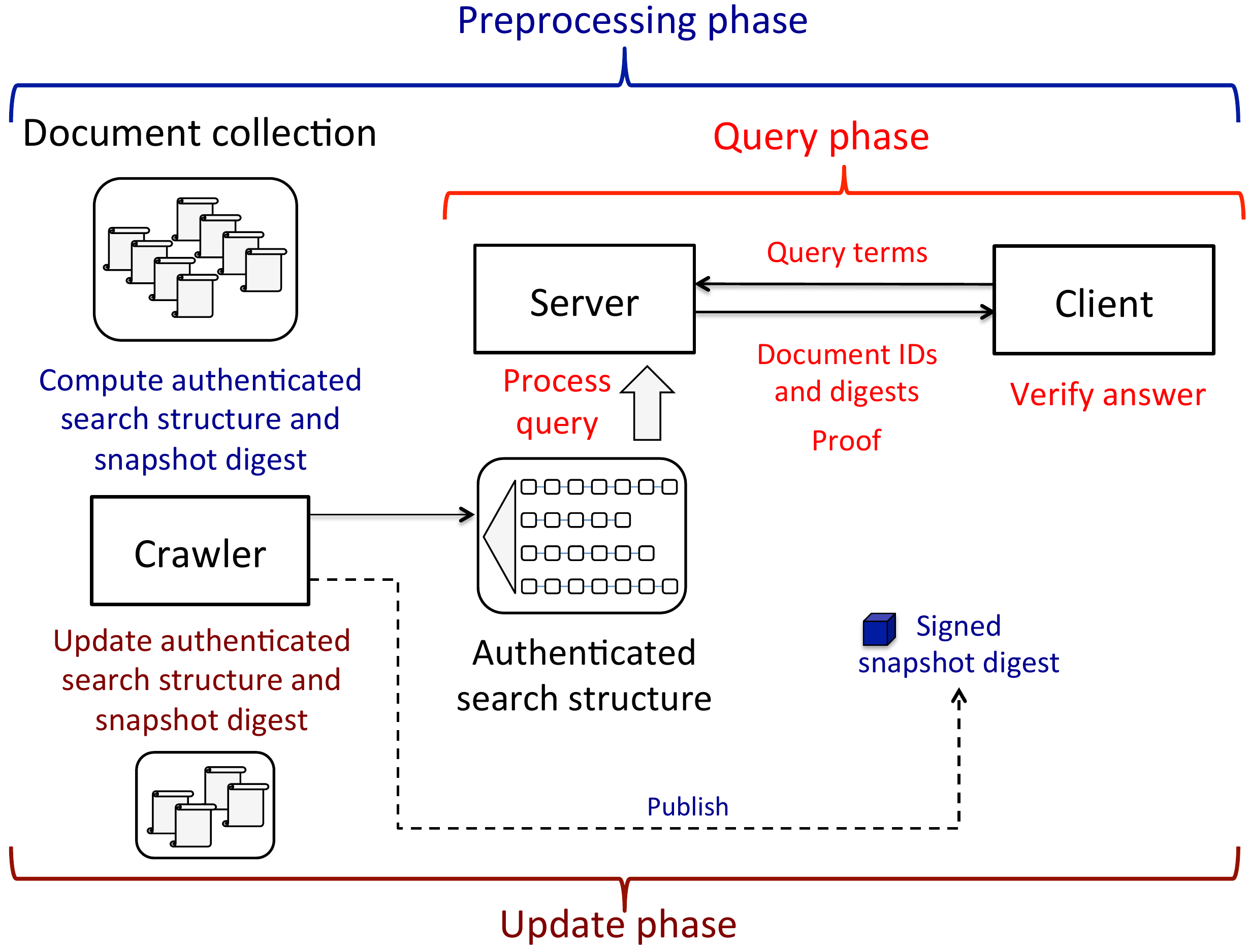}
\caption{\label{crawl}The three-party model as the main operational
  setting of an authenticated web crawler.}
\end{figure}
\fi

We refer to the three-party model of Figure~\ref{crawl}. (We note that
although relevant, this model is considerably different than the
``traditional'' three-party model that has been studied in the field of
authenticated data structures\ifFull ~\cite{t-ads-03}\fi.) In our model
the three parties are called \emph{crawler}, \emph{server}, and
\emph{client}. The client trusts the crawler but not the server.  The
protocol executed by the parties consists of three phases, preprocessing
phase, query phase and update phase.

\ifShort
\begin{figure}
\centering
\includegraphics[scale=0.335]{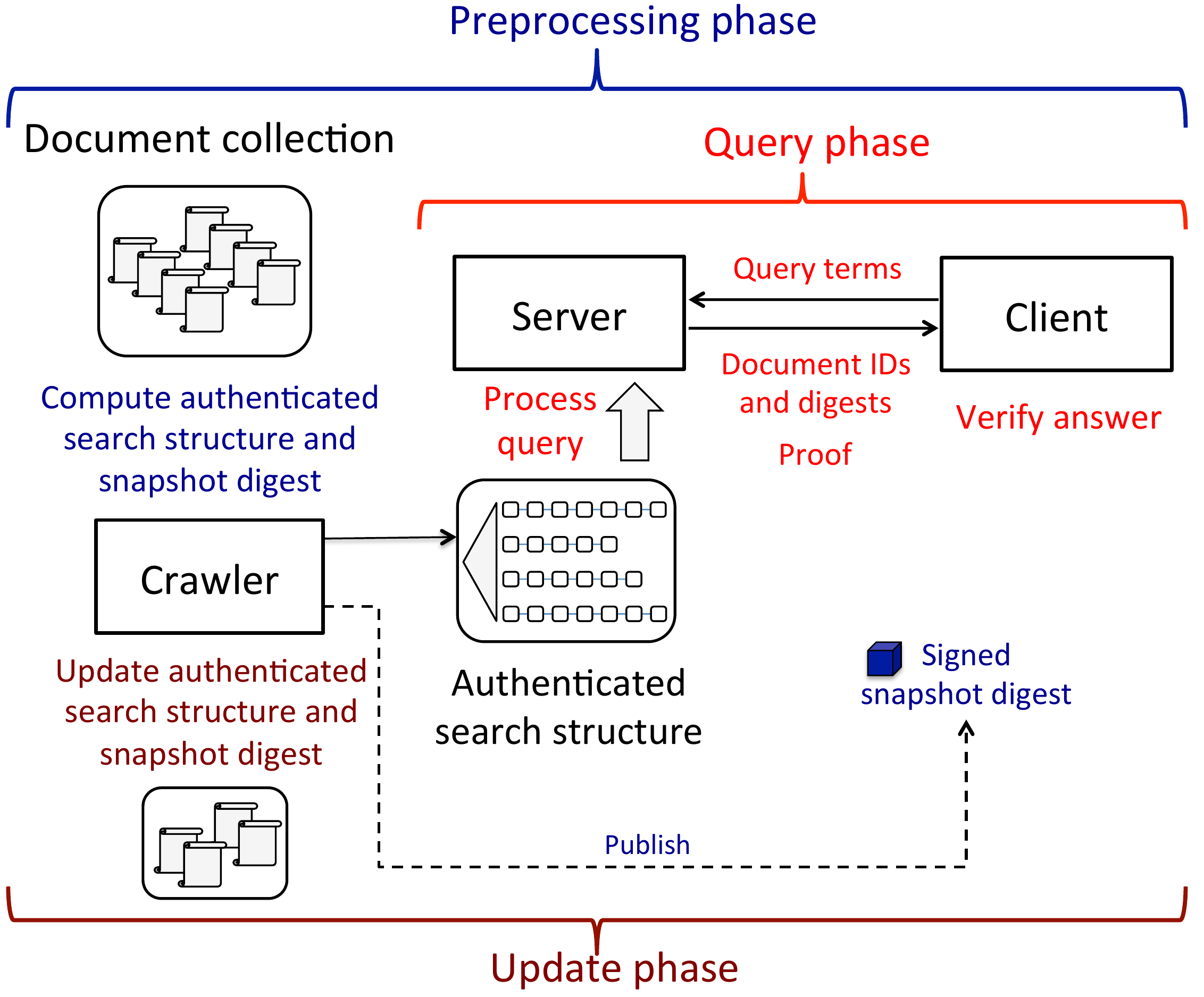}
\caption{\label{crawl}The three-party model as the main operational
  setting of an authenticated web crawler.}
\end{figure}
\fi

\ifFull
\begin{figure}[t]
\centering
\includegraphics[scale=0.6]{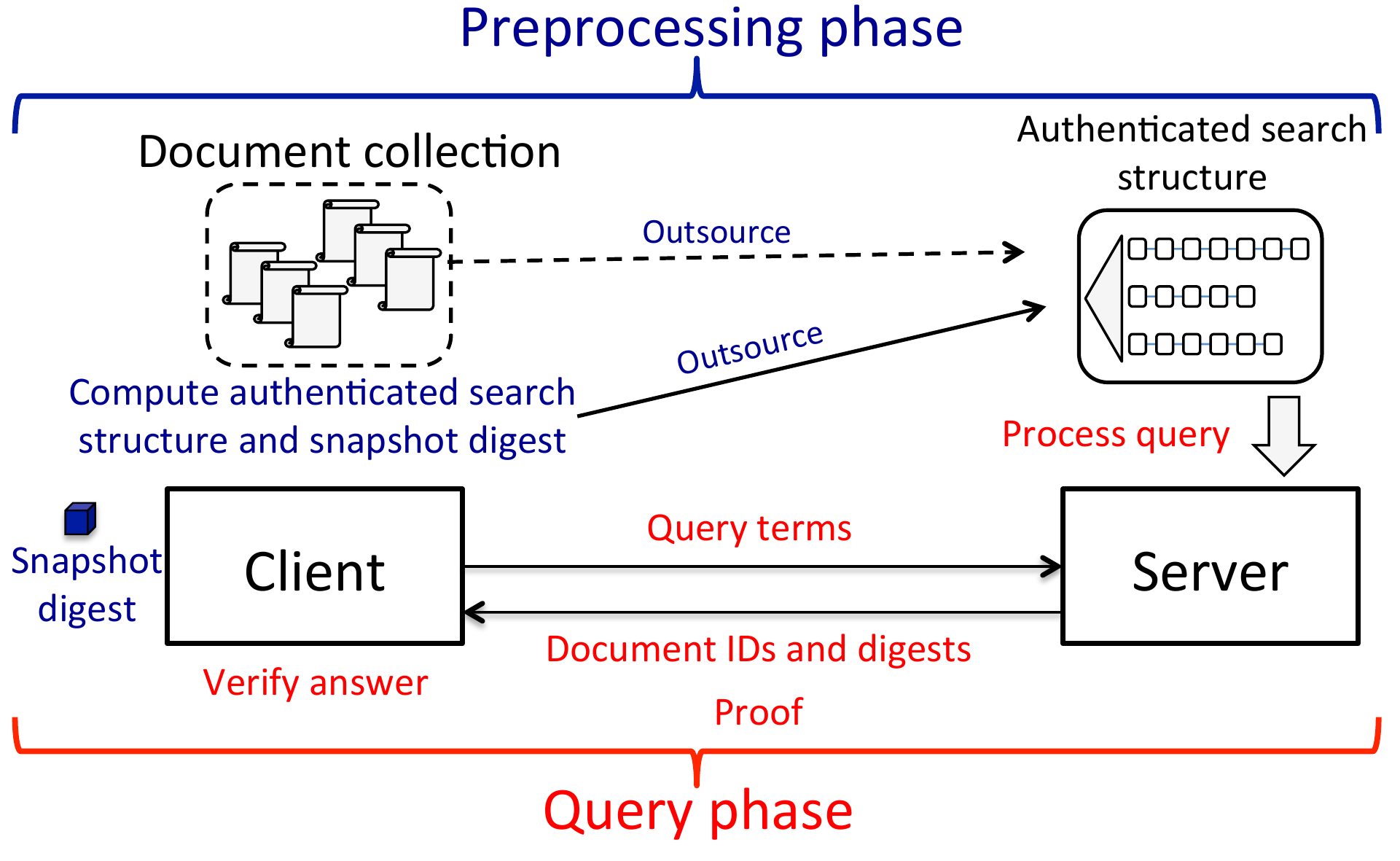}
\caption{\label{fig:crawl_twoparty1} The first two-party model: Client
  outsources a document collection and uses server to store it and
  perform search queries on it.}
\end{figure}
\begin{figure}[t]
\centering
\includegraphics[scale=0.6]{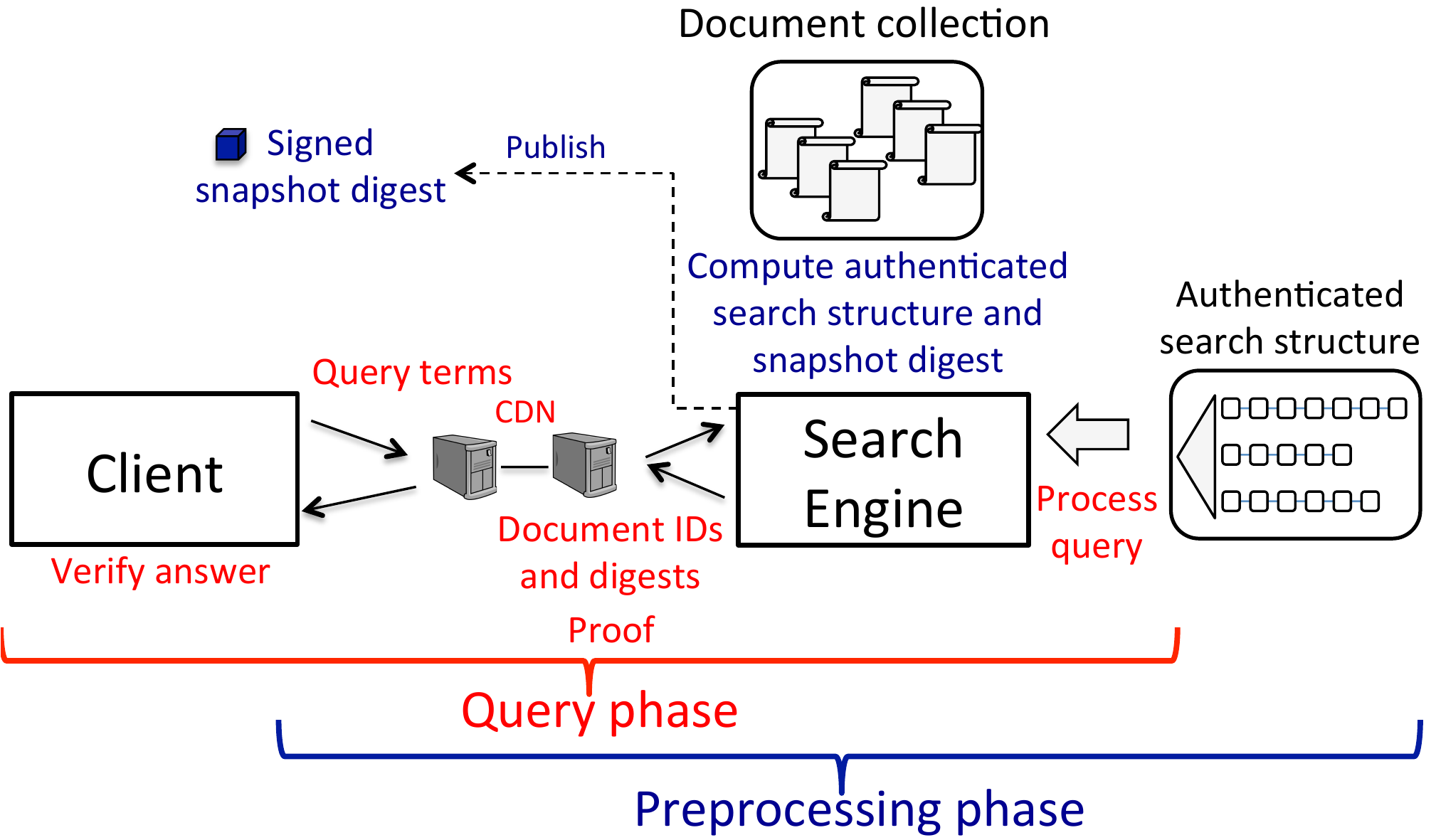}
\caption{\label{fig:crawl_twoparty2}The second two-party model: Search
  engine protects his results from the man-in-the-middle attack.}
\end{figure}
\fi

In the \emph{preprocessing phase}, the crawler accesses the collection
of documents and takes a snapshot of them.  The crawler then produces
and digitally signs a secure, collision-resistant digest of this
snapshot. This digest contains information about both the identifiers of
the documents in the snapshot and their contents.  The signed digest is
made public so that clients can use it for verification during the query
phase.  Concurrently, the crawler builds an authenticated data structure
supporting keyword searches on the documents in the snapshot.
Finally, the crawler outsources to the server the snapshot of the
collection along with the authenticated data structure.

In the \emph{query phase}, the client sends a query request, consisting
of keywords, to the server. The server computes the query results, i.e.,
the set of documents in the snapshot that contain the query
keywords. Next, the server returns to the client an answer consisting of
the computed query results and
a cryptographic proof that the query results are correct, complete,
fresh\footnote{We note that in the cryptographic literature on
  authenticated data structures (e.g.,~\cite{ptt-aht-08,ptt-ovods-11})
  these three integrity properties are combined into a unified
  \emph{security} property.}  and that their associated digests are
valid. The client verifies the correctness and the completeness of the
query results using the proof provided by the server and relying on the
trust in the snapshot digest that was signed and published by the
crawler.
The above interaction can be repeated for another query issued by the
client.

In the \emph{update phase}, the crawler parses the documents to be added
or removed from the collection and sends to the server any new contents
and changes for the authenticated data structure.  Additionally, it
computes and publishes a signed digest of the new snapshot of the
collection.

Note that the query phase is executed after the preprocessing phase.
Thus, the system gives assurance to the client that the documents have
not been changed since the last snapshot was taken by the crawler.
Also, during the update phase, the newly signed digest prevents the
server from tampering with the changes it receives from the crawler.

\subsection{Two-party Models}
\label{sec:other_models}
Our solution can also be used in a two-party model in two scenarios.  In
Figure~\ref{fig:crawl_twoparty1}, we consider a cloud-based model where
a client outsources its data along with an authenticated search
structure to the cloud-based server but keeps a digest of the snapshot
of her data.  In this scenario, the server executes keyword-search
queries issued by the client and constructs a proof of the result.
The client can then verify the results using the digest kept before
outsourcing.  The model shown in Figure~\ref{fig:crawl_twoparty2}
protects the interaction between a client and a search engine from a
man-in-the-middle attack. The search engine publishes a signed digest of
its current snapshot of the web and supplies every search result with a
proof.  Upon receiving search results in response to a query, the client
can use the digest and the proof to verify that the results sent from
the search engine were not tampered with.

\subsection{Desired Properties}
The main properties we seek to achieve in our solution are
\emph{security} and \emph{efficiency}. The system should be secure,
i.e., it should be computationally infeasible for the server to
construct a verifiable proof for an incorrect result (e.g., include a
web page that does not contain a keyword of the query). Our system
should also be practical, i.e., we should avoid a solution where the
authenticated crawler literally signs every input participating in the
specific computation and where the verification is performed by
downloading and verifying all these inputs.  For example, consider a
search for documents which contain two specific terms. Each term could
appear in many documents while only a small subset of them may contain
both. In this case, we would not want a user to know about all the
documents where these terms appear in order to verify that the small
intersection she received as a query answer is correct.
Finally, we want our solution to support efficient updates: Due to their
high frequency, updates on web contents should incur overhead that is
proportional to the size of the updated content and not of the entire
collection.

We finally envision two modes of operations for our authentication
framework. First, the authenticated web crawler operates autonomously;
here, it serves as a ``web police officer'' and creates a global
snapshot of a web site. This mode allows the integrity checking to be
used as a verification that certain access control or compliance
policies are met (e.g., a page owner does not post confidential
information outside its organizational domain). Alternatively, the
authenticated web crawler operates in coordination with the web page
owners (authors); here, each owner interacts with the crawler to commit
to the current local snapshot of its web page. This mode allows the
integrity proofs to be transferable to third parties in case of a
dispute (e.g., no one can accuse a page owner for deliberately posting
offending materials).



\section{Our solution}
\label{sec:solution}
In this section we describe the cryptographic and algorithmic methods
used by all the separate entities in our model for the verification of
conjunctive keyword searches. As we will see, conjunctive keyword
searches are equivalent with a set intersection on the underlying
\emph{inverted index data structure}~\cite{zm-iftse-06}. Our solution is
based on the authenticated data structure of~\cite{ptt-ovods-11} for
verifying set operations on outsourced sets. We now provide an overview
of the construction in~\cite{ptt-ovods-11}. First we introduce the
necessary cryptographic primitives.

\subsection{Cryptographic Background} 

Our construction makes use of the cryptographic primitives of
\emph{bilinear pairings} and \emph{Merkle hash trees}.

\smallskip\noindent{\bf Bilinear pairings.} Let $\mathbb{G}$ be a cyclic
multiplicative group of prime order $p$, generated by $g$. Let also
$\mathcal{G}$ be a cyclic multiplicative group with the same order $p$
and $e:\mathbb{G}\times \mathbb{G}\rightarrow \mathcal{G}$ be a bilinear
pairing with the following properties: (1) Bilinearity:
$e(P^a,Q^b)=e(P,Q)^{ab}$ for all $P,Q\in \mathbb{G}$ and $a,b\in
\mathbb{Z}_p$; (2) Non-degeneracy: $e(g,g)\ne 1$; (3) Computability:
There is an efficient algorithm to compute $e(P,Q)$ for all $P,Q\in
\mathbb{G}$. We denote with $(p,\mathbb{G},\mathcal{G},e,g)$ the
bilinear pairings parameters, output by a polynomial-time algorithm on
input~$1^k$.

\smallskip\noindent{\bf Merkle hash trees.} A Merkle hash
tree~\cite{m-cds-89} is an authenticated data structure
\ifFull\cite{t-ads-03} \fi allowing an untrusted server to vouch for the
integrity of a dynamic set of indexed data $\T[0],\T[1],\ldots,\T[m-1]$
that is stored at untrusted repositories. Its representation comprises a
binary tree with hashes in the internal nodes, denoted with
$\mathsf{merkle}(\T)$. A Merkle hash tree is equipped with the following
algorithms:
\begin{enumerate}
\item
  $\{\mathsf{merkle}(\T),\mathsf{sig}(\T)\}=\textsf{setup}(\T)$. This
  algorithm outputs a succinct signature of the table $\T$ which can be
  used for verification purposes and the hash tree
  $\mathsf{merkle}(\T)$.  To construct the signature and the Merkle
  tree, a collision-resistant hash function $\textsf{hash}(\cdot)$ is
  applied recursively over the nodes of a binary tree on top of $\T$.
  Leaf $\ell\in\{0,1,\ldots,m-1\}$ of $\mathsf{merkle}(\T)$ is assigned
  the value $h_{\ell}=\mathsf{hash}(\ell||\T[\ell])$, while each
  internal node $v$ with children $a$ and $b$ is assigned the value
  $h_v=\mathsf{hash}(h_a||h_b)$. The root of the tree $h_r$ is signed to
  produce signature~$\mathsf{sig}(\T)$.

\item $\{\mathsf{proof}(i),\mathsf{answer}(i)\}=
  \textsf{query}(i,\T,\mathsf{merkle}(\T))$. Given an index $0\le i\le
  m-1$, this algorithm outputs a proof that could be used to prove that
  $\mathsf{answer}(i)$ is the value stored at $\T[i]$.  Let
  $\mathsf{path}(i)$ be a list of nodes that denotes the path from leaf
  $i$ to the root and $\mathsf{sibl}(v)$ denote a sibling of node $v$ in
  $\mathsf{merkle}(\T)$.  Then, $\mathsf{proof}(i)$ is the ordered list
  containing the hashes of the siblings $\mathsf{sib}(v)$ of the nodes
  $v$ in $\mathsf{path}(i)$.

\item $\{0,1\}=\textsf{verify}(\mathsf{proof}(i),
  \mathsf{answer}(i),\mathsf{sig}(\T))$. This algorithm is used for
  verification of the answer $\mathsf{answer}(i)$. It computes the root
  value of $\mathsf{merkle}(\T)$ using $\mathsf{answer}(i)$ and
  $\mathsf{proof}(i)$, i.e., the sibling nodes of nodes in
  $\mathsf{path}(i)$, by performing a chain of hash computations over
  nodes in $\mathsf{path}(i)$. It then checks to see if the output is
  equal to $h_r$, a value signed with $\mathsf{sig}(\T)$, in which case
  it outputs $1$, implying that $\mathsf{answer}(i)=\T[i]$ whp.
\end{enumerate}
\ifFull
For more details on Merkle hash trees, please refer to~\cite{m-cds-89}.
\fi

\ifFull
\begin{figure}[t]
\centering
\ifShort
\includegraphics[scale=0.6]{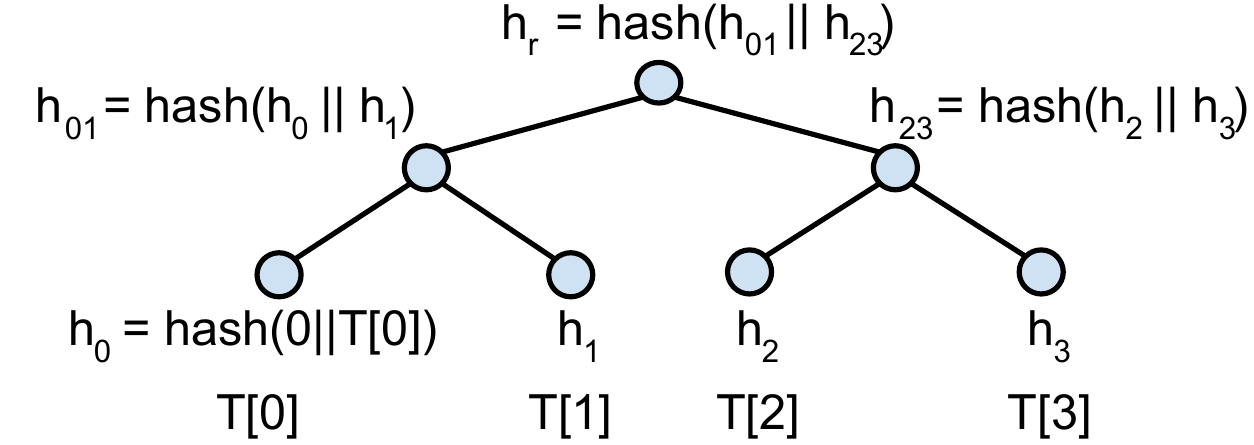}
\else
\includegraphics[scale=0.7]{example_merkle_tree0}
\fi
\ifShort
\vspace{-3mm}
\fi
\caption{Example of a Merkle hash tree.}
\label{fig:merkle_tree}
\ifShort
\vspace{-5mm}
\fi
\end{figure}
\fi

\ifFull
\smallskip\noindent \emph{Example:} Consider a Merkle hash tree for four
data items in Figure~\ref{fig:merkle_tree}.  A collision resistant hash
function $\mathsf{hash}$ is used recursively to compute a hash value for
each node of the tree.  On $\textsf{query}(2,\T,\mathsf{merkle}(\T))$
the query algorithm returns
\begin{eqnarray*}
  \mathsf{answer}(2) &=& \T[2]\,, \\
  \mathsf{proof}(2) &=& \{h_3, h_{01}\}\,.
\end{eqnarray*}
The verification algorithm then computes the values $h_{2'} =
\mathsf{hash}(2||\T[2])$, $h_{23'} = \mathsf{hash}(h_{2'}||h_3)$, and then
checks that
\[
h_r \stackrel{?}{=} \mathsf{hash}(h_{01}||h_{23'})\,.
\]
\fi 

We now describe in detail the individual functionality of the \emph{web
  crawler}, the \emph{untrusted server} and the \emph{client}.

\subsection{Web Crawler}
The web page collection we are interested in verifying is described with
an inverted index data structure. Suppose there are $m$ terms
$q_0,q_2,\ldots,q_{m-1}$ over which we are indexing, each one mapped to
a set of web pages $S_i$, such that each web page in $S_i$ contains the
term $q_i$, for $i=0,1,\ldots,m-1$. Assume, without loss of generality,
that each web page in $S_i$ can be represented with an integer in
$\mathbb{Z}^{*}_p$, where $p$ is large $k$-bit prime. For example, this
integer can be is a cryptographic hash of the entire web page. However,
if we want to cover only a subset of the data in the page, the hash
could be applied to cover text and outgoing links, but not the HTML
structure. The extraction of the relevant data will be made by a special
filter and will depend on the application. For instance, if we are
interested in tables of NYSE financial data, there is no need to include
images in the hash.


The authenticated data structure is built very simply as follows: First
the authenticated crawler picks \emph{random value} $s\in \mathbb{Z}^{*}_p$
which is kept secret. Then, for each set $S_i$ ($i=0,1,\ldots,m-1$), the
\emph{accumulation value},
$$\T[i]=g^{\prod_{x\in S_i}(s+x)}\,,$$ 
is computed, where $g$ is a generator of the group $\mathbb{G}$ from an
instance of bilinear pairing parameters. Then the crawler calls
algorithm $\textsf{setup}(\T)$ to compute signature $\mathsf{sig}(\T)$
and Merkle hash tree $\mathsf{merkle}(\T)$; the former is passed to the
clients (to support the result verification) and the latter is
outsourced to the server (to support the proof computation).

\smallskip
\noindent{\bf Intuition.}
There are \emph{two} integrity-protection levels: First, the Merkle hash
tree protects the integrity of the accumulation values $\T[i]$ offering
coarse-grained verification of the integrity of the sets. Second, the
accumulation value $\T[i]$ of $S_i$ protects the integrity of the web
pages in set $S_i$ offering fine-grained verification of the sets. In
particular, each accumulation value $\T[i]$ maintains an algebraic
structure within the set $S_i$ that is useful in two
ways~\cite{ptt-aht-08,ptt-ovods-11}: (1) subject to an authentic
accumulation value $\T[i]$ (subset) membership can be proved using a
succinct witness, yielding a proof of ``correctness,'' and (2)
disjointness between $t$ such sets can be proved using $t$ succinct
witnesses, yielding a proof of ``completeness.'' Bilinearity is crucial
for testing both properties.

\smallskip
\noindent {\bf Handling updates.}
During an update to inverted list $S_i$ of term $q_i$ the crawler needs
to change the corresponding accumulation value $\T[i]$ and update the
path of the Merkle tree from the leaf corresponding to term $q_i$ to the
root.  A new page $x'$ is added to an inverted list $S_i$ of term $q_i$
if either a new page $x'$ contains term $q_i$ or the content of page
$x'$ that is already in the corpus is updated and now contains $q_i$.
In this case the accumulation value of $q_i$ is changed to $\T'[i] =
\T[i]^{(s+x')}$.  If some web page $x'\in S_i$ is removed or no longer
contains term $x'$ the accumulation value is changed to $\T'[i] =
\T[i]^{1/(s+x')}$.  It is straightforward to handle updates in the
Merkle hash tree (in time logarithmic in the number of inverted
lists---see also Section~\ref{sec:performance}).

\subsection{Untrusted Server}
\label{sec:solution_untrusted_server}
The untrusted server in our solution stores the inverted index along
with the authentication information defined earlier as authenticated
data structure $\mathsf{merkle}(\T)$. Given a conjunctive keyword-search
query $q=(q_1,q_2,\ldots,q_t)$ from a client, the server returns a set of
web pages $\I$ where each web page $p\in \I$ contains all terms from
$q$. Namely, it is the case that \ifFull
\[
\I=S_1\cap S_2\cap\ldots \cap S_t\,.
\]
\fi The server is now going to
compute a proof so that a client can verify that all web pages included
in $\I$ contain $q_1,q_2,\ldots,q_t$ and ensure that no web page from
the collection that satisfies query $q$ is omitted from $\I$. Namely the
server needs to prove that $\I$ is the correct intersection $S_1\cap
S_2\cap\ldots \cap S_t$.

One way to compute such a proof would be to have the server just send
\emph{all} the elements in $S_1,S_2,\ldots,S_t$ along with Merkle tree
proofs for $\T[1],\T[2],\ldots,\T[t]$. The contents of these sets could
be verified and the client could then compute the intersection
locally. Subsequently the client could check to see if the returned
intersection is correct or not. The drawback of this solution is that it
involves \emph{linear communication and verification complexity}, which
could be prohibitive, especially when the sets are large.

To address this problem, in CRYPTO 2011, Papamanthou, Tamassia and
Triandopoulos~\cite{ptt-ovods-11} observed that it suffices to certify
\emph{succinct relations} related to the correctness of the
intersection. These relations have size independent of the sizes of the
sets involved in the computation of the intersection, yielding a very
efficient protocol for checking the correctness of the
intersection. Namely $\I=S_1\cap S_2\cap\ldots \cap S_t$ is the correct
intersection if and only if:
\begin{enumerate}
\item $\I \subseteq S_1 \wedge \ldots \wedge \I \subseteq S_t$ (subset
  condition);
\item $(S_1 - \I) \cap \ldots \cap (S_t - \I) = \emptyset$ (completeness
  condition).
\end{enumerate}

Accordingly, for every intersection $\I = \{y_1,y_2,\ldots,y_{\delta}\}$
the server constructs the proof that consists of four parts:
\begin{description}
\item[A] Coefficients $b_\delta, b_{\delta-1},\ldots, b_0$ of the
  polynomial $(s+y_1)(s+y_2) \ldots (s+y_\delta)$ associated with the
  intersection $\I$;
\item[B] Accumulation values $\T[j]$ associated with the sets $S_j$, along
  with their respective proofs $\mathsf{proof}(j)$, output from calling
  algorithm $\mathsf{query}(j, \T,\mathsf{merkle}(\T))$, for
  $j=1,\ldots,t$;
\item[C] Subset witnesses $\W_{\I,j}=g^{P_j(s)}$, for $j = 1,\ldots, t$,
  where
\[
P_j(s)=\prod_{x\in S_j-\I}(x+s)\,;
\]
\item[D] Completeness witnesses $\F_{\I,j} = g^{q_j(s)}$ for $j =
  1,\ldots, t$, such that
  \ifFull
\[
q_1(s)P_1(s) + q_2(s)P_2(s) + \ldots + q_t(s)P_t(s) =1\,,
\] 
\fi
where $P_j(s)$ are the exponents of the subset witnesses.
\end{description}

\smallskip
\noindent{\bf Intuition.} Part A comprises an encoding of the result (as
a polynomial) that allows efficient verification of the two
conditions. Part B comprises the proofs for the 1-level integrity
protection based on Merkle hash trees.  Part C comprises a
subset-membership proof for the 2-level integrity protection based on
the bilinear accumulators.  Part D comprises a set-disjointness proof
for the 2-level integrity protection based on the extended Euclidean
algorithm for finding the interrelation of irreducible polynomials.

\subsection{Client} 
\label{client_verify}
The client verifies the intersection $\I$ by appropriately verifying the
corresponding proof elements described above:
\begin{description}

\item[A] It first certifies that coefficients $b_\delta, b_{\delta-1},
  . . . , b_0$ are computed correctly by the server, i.e., that they
  correspond to the polynomial $\prod_{x\in \I}(s+x)$, by checking that
  $\sum_{i=0}^{\delta}b_i\kappa^i$ equals $\prod_{x\in \I}(\kappa+x)$
  for a randomly chosen value $\kappa\in \mathbb{Z}^*_p$;

\item[B] It then verifies $\T[j]$ for each term $q_j$ that belongs to
  the query ($j=1,\ldots,t$), by using algorithm
  $\textsf{verify}(\mathsf{proof}(j),$ $\T[j],\mathsf{sig}(\T))$;

\item[C] It then checks the subset condition
\[
e\left(\prod_{k=0}^\delta(g^{s^k})^{b_k}, \W_{\I,j}\right) = e\left(\T[j], g\right) \text{ for } j=1,\dots, t\,;
\]

\item[D] Finally, it checks that the completeness condition holds
\begin{equation}\label{complete_condition}
\prod_{j=1}^{t} e\left(\W_{\I,j}, \F_{\I, j}\right) = e(g,g)\,.
\end{equation}
\end{description}
The client accepts the intersection as correct if and only if all the
above checks succeed.

\smallskip
\noindent{\bf Intuition.} Step A corresponds to an efficient,
high-assurance probabilistic check of the consistency of the result's
encoding. Steps C and D verify the correctness of the subset and
set-disjointness proofs based on the bilinearity of the underlying group
and cryptographic hardness assumptions that are related to discrete-log
problems.

\subsection{Final Protocol}
We now summarize the protocol of our solution.

\smallskip\noindent \textbf{Web crawler.} Given a security parameter:
\begin{enumerate}
\item Process a collection of webpages and create an inverted index.
\item Generate a description of the group and bilinear pairing
  parameters $(p,\mathbb{G},\mathcal{G},e,g)$.
\item Pick randomly a secret key $s\in \mathbb{Z}^{*}_p$.
\item Compute accumulation value $\T[i]$ for each term $i$ in the
  inverted index.
\item Build a Merkle hash tree $\mathsf{merkle}(\T)$ and sign the root
  of the tree $h_r$ as $\mathsf{sig}(\T)$.
\item Compute values $g^{s^1},\ldots, g^{s^n}$, where \\ $n \ge \max\{m,
  \max_{i=0,...,m-1}\{|S_i|\}\}$.
\item Send inverted index and $\mathsf{merkle}(\T)$ to the server.
\item Publish $\mathsf{sig}(\T)$, $(p,\mathbb{G},\mathcal{G},e,g)$ and
  $g^{s^1},g^{s^2},\ldots,g^{s^n}$ so that the server can access them to
  compute the proof and clients can acquire them during verification.
\end{enumerate}

\smallskip\noindent \textbf{Untrusted server.} Given a query $q= \{q_1, q_2,
\ldots, q_t\}$:
\begin{enumerate}
\item Compute the answer for $q$ as the intersection $\I =
  \{y_1,y_2,\ldots,y_{\delta}\}$ of inverted lists corresponding to $q$.
\item Compute the coefficients $b_\delta, b_{\delta-1}, \ldots, b_0$
  corresponding to the polynomial
  $(s+y_1)(s+y_2),\ldots,(s+y_{\delta})$.
\item Use $\mathsf{merkle}(\T)$ to compute the integrity proofs of
  $\T[j]$.
\item Compute subset witnesses $\W_{\I,j}=g^{P_j(s)}$.
\item Compute completeness witnesses $\F_{\I,j}=g^{q_j(s)}$.
\item Send $\I$ and all components of the proof to the client.
\end{enumerate}

\smallskip\noindent {\bf Client.} Send query $q$ to the server, and
given an answer to the query and the proof, accept the answer as correct
if all of the following hold:
\begin{enumerate}
\item Coefficients of the intersection are computed correctly: Pick a
  random $\kappa \in \mathbb{Z}^{*}_p$ and verify that
  $\sum_{k=0}^{\delta} b_i \kappa^i = \prod_{x\in \I} (\kappa + x)$.
  (Note that the client can verify the coefficients without knowing the
  secret key~$s$.)
\item Accumulation values are correct: Verify integrity of these values
  using $\mathsf{sig}(\T)$ and $\mathsf{merkle}(\T)$.
\item Subset and completeness conditions hold.
\end{enumerate}

\begin{figure}[t!]
\centering
\includegraphics[width=78mm]{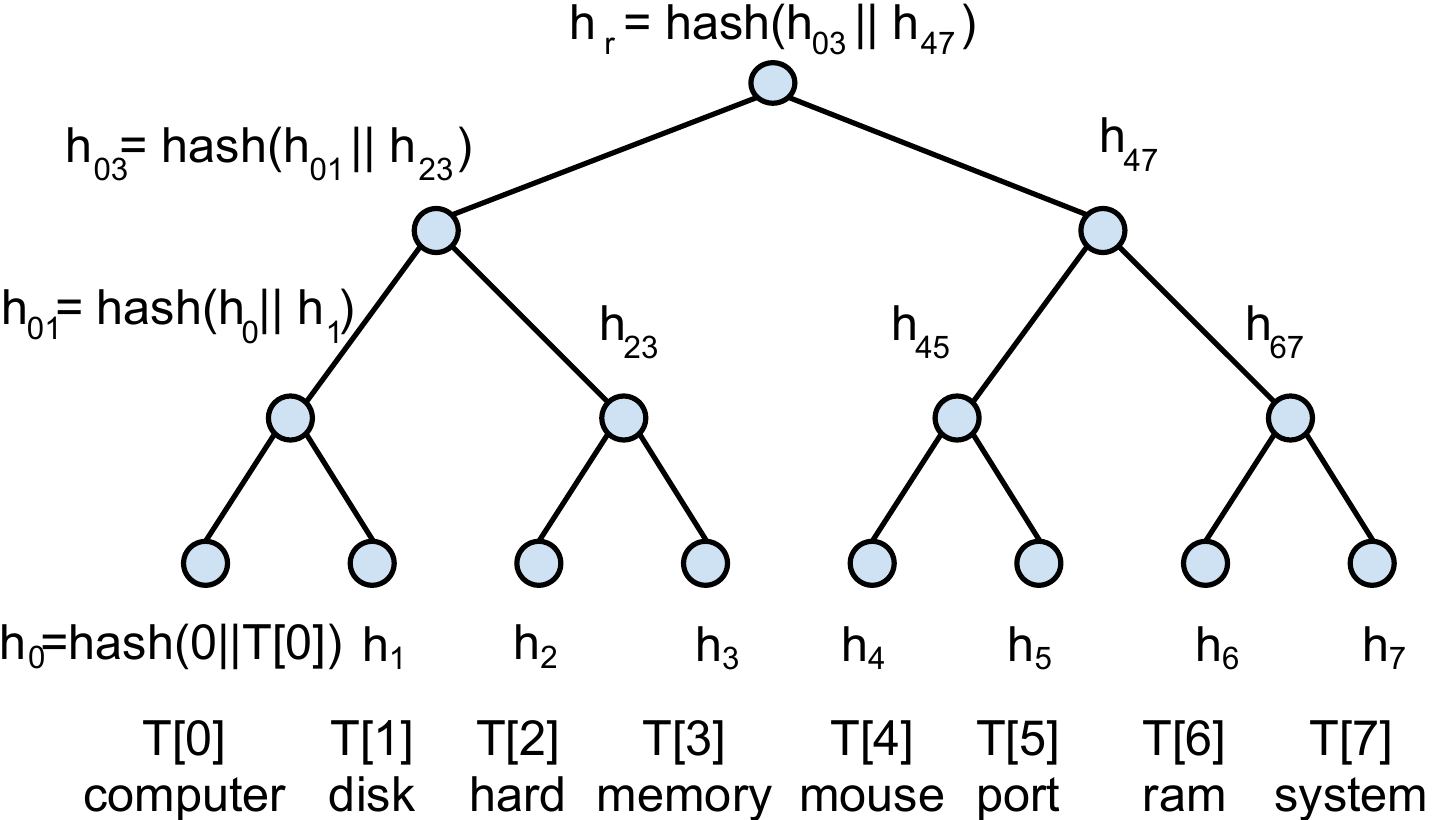}
\caption{Merkle tree for authenticating the accumulation values of the
  terms in Table~\ref{tbl:example_inv_list}.}
\label{fig:example_merkle_tree}
\end{figure}

\noindent
\begin{table}[h!]
\begin{center}
\begin{tabular}{c|r|l}
Term ID & Term & Inverted list \\
\hline
0 & computer & 6,8,9 \\
1 & disk & 1,2,4,5,6,7 \\
2 & hard & 1,3,5,7,8,9 \\
3 & memory & 1,4,7 \\
4 & mouse & 2,5 \\
5 & port & 3,5,9 \\
6 & ram & 5,6,7 \\
7 & system & 1,7 \\
\end{tabular}
\end{center}
\caption{An example of an inverted index.}
\label{tbl:example_inv_list}
\end{table}%
\noindent \emph{Example:} We now consider how our protocol works on a
toy collection.  Consider an inverted index in
Table~\ref{tbl:example_inv_list} where a term is mapped to a set of
documents where it appears, i.e., an inverted list.  For example, term
``mouse'' appears in documents 2 and 5 and document 2 contains words
``disk'' and ``mouse''.  The crawler computes accumulation values
$\T[i]$ for each term id $i$, e.g., an accumulation value for term
``memory'' is \ifFull
\[
\T[3] = g^{(s + 1)(s + 4)(s + 7)},
\]
and builds a Merkle Tree where each leaf corresponds to a term, see
Figure~\ref{fig:example_merkle_tree}.  
\fi 
The Merkle tree and the inverted index are sent to the server.  The
crawler also computes $g^{s^1},g^{s^2}, \ldots, g^{s^8}$ and publishes
them along with a signed root of the Merkle tree, $\mathsf{sig}(\T)$.

Given a query $q=(\emph{hard}\ \textsf{AND}\ \emph{disk}\ \textsf{AND}\
\emph{memory})$, the result is the intersection $\I$ of the inverted
lists for each of the terms in the query. In our case, $\I=\{1,7\}$.
The server builds a proof and sends it to the client along with the
intersection.  The proof consists of the following parts:

\begin{description}

\item[A] Intersection proof: Coefficients $b_0=7$, $b_1=8$ and $b_2=1$
  of the intersection polynomial $(s+1)(s+7)$.

\item[B] \ifFull Accumulation values proof: values $\T[0]$, $\T[1]$ and
  $\T[2]$ with a proof from the Merkle tree that these values are
  correct:
\begin{eqnarray*}
\T[1]: \{{h_0}, {h_{23}}, {h_{47}}\}\\
\T[2]: \{{h_3}, {h_{01}}, {h_{47}}\}\\
\T[3]: \{{h_2}, {h_{01}}, {h_{47}}\}\\
\end{eqnarray*}
\fi

\item[C] Subset witnesses $g^{P_1(s)}$, $g^{P_2(s)}$ and $g^{P_3(s)}$
  for each term in the query where
\begin{eqnarray*}
P_1(s) &=& (s+2)(s+4)(s+5)(s+6)\,, \\
P_2(s) &=& (s+3)(s+5)(s+8)(s+9)\,, \\
P_3(s) &=& (s+4)\,.
\end{eqnarray*}

\item[D] Completeness witnesses: Using the Extended Euclidean algorithm
  the server finds values $g^{q_1(s)}$, $g^{q_2(s)}$ and $g^{q_3(s)}$
  such that 
\ifFull
\[
q_1(s)P_1(s) + q_2(s)P_2(s) + q_3(s)P_3(s)  = 1
\]
\fi

\end{description}
Note that since the server knows values $g^{s^i}$, it can compute values $g^{P_j(s)}$ and $g^{q_j(s)}$ without knowing a private key~$s$ (only the coefficients of the polynomials are required).

To verify the response from the server, the client:

\begin{description}
\item[A] Picks random $\kappa \in \mathbb{Z}^{*}_p$ and checks that
  $\sum_{k=0}^{\delta} b_i \kappa^i =$\\
  $\prod_{x\in \I} (\kappa + x)$, e.g., $7 + 8\kappa +\kappa^2 =
  (\kappa+1) (\kappa+7)$.
\item[B] Verifies that each $\T[i]$ is correct, e.g., for $\T[1]$ and
  its proof $\{{h_0}$, ${h_{23}}$, ${h_{47}}\}$ the client checks that
\[
h_r \stackrel{?}{=} \textsf{hash}(\textsf{hash}(\textsf{hash}({h_0},\textsf{hash}(1||\T[1])), {h_{23}}), {h_{47}}),
\]
such that $\mathsf{sig}(\T)$ is a signed root of the Merkle tree $h_r$. 
\item[C] Checks subset condition:
\[
e\left(\prod_{k=0}^2(g^{s^k})^{b_k}, g^{P_j(s)}\right) = e\left(\T[j], g\right) \text{ for } j=1,2\,.
\]
\item[D] Checks completeness: 
\ifFull
\[
\prod_{j=1}^{3} e\left(g^{P_j}(s), g^{q_j(s)}\right) = e(g,g)\,.
\]
\fi
\end{description}

\subsection{Security}
With respect to the security, we show that given an intersection query
(i.e., a keyword search) referring to keywords $q_1,q_2,\ldots,q_t$, any
computationally-bounded adversary cannot construct an incorrect answer
$\mathcal{I}$ and a proof $\pi$ that passes the verification test of
Section~\ref{client_verify}, except with negligible probability.  The
proof is by contradiction. Suppose the adversary picks a set of indices
$q=\{1,2,\ldots,t\}$ (wlog), all between $1$ and $m$ and outputs a proof
$\pi$ and an \emph{incorrect} answer $\mathcal{I}\ne
\mathsf{I}=S_{1}\cap S_{2}\cap \ldots \cap S_{t}$. Suppose the answer
$\alpha(q)$ contains $d$ elements. The proof $\pi$ contains (i)
\emph{Some} coefficients $b_0,b_1,\ldots,b_d$; (ii) \emph{Some}
accumulation values $\mathsf{acc}_j$ with \emph{some} respective proofs
$\Pi_j$, for $j=1,\ldots,t$; (iii) \emph{Some} subset witnesses
$\mathsf{W}_j$ with \emph{some} completeness witnesses $\mathsf{F}_j$,
for $j=1,\ldots,t$ (inputs to the verification algorithm).

Suppose the verification test on these values is successful. Then: (a)
By the certification procedure, $\beta_0,\beta_1,\ldots,\beta_d$ are
\emph{indeed} the coefficients of the polynomial $\prod_{x\in
  \mathcal{I}}(x+s)$, except with negligible probability; (b) By the
properties of the Merkle tree, values $\mathsf{acc}_j$ are \emph{indeed}
the accumulation values of sets $S_j$, except with negligible
probability; (c) By the successful checking of the subset condition,
values $\mathsf{W}_{j}$ are \emph{indeed} the subset witnesses for set
$\mathcal{I}$ (with reference to $S_j$), i.e.,
$\mathsf{W}_j=g^{P_j(s)}$, except with negligible probability; (d)
However, since $\mathcal{I}$ is incorrect then it cannot include
\emph{all} the elements and there must be at least one element $a$ that
is not in $\mathcal{I}$ and is a \emph{common factor} of polynomials
$P_1(s),P_2(s),\ldots,P_t(s)$.  In this case, the adversary can divide
the polynomials $P_1(s),P_2(s),\ldots,P_t(s)$ with $s+a$ in the
completeness relation of Equation~\ref{complete_condition} and derive
the quantity $e(g,g)^{1/(s+a)}$ at the right hand side. However, this
implies that the adversary has solved in polynomial time a difficult
problem in the target group $\mathcal{G}$,
in particular, the adversary has broken the bilinear $q$-strong
Diffie-Hellman assumption, which happens with negligible
probability. More details on the security proof can be found
in~\cite{ptt-ovods-11}.
%

\section{Performance}
\label{sec:performance}

In this section, we describe a prototype implementation of our
authenticated crawler system and discuss the results of experimentation
regarding its practical performance.

\subsection{Performance Measures}
\label{sec:asymp_comp}
We are interested in studying four performance measures:
\begin{enumerate}

\item The size of the proof of a query result, which is sent by the
  server to the client.  Asymptotically, the proof size is $O(\delta +
  t\log m)$ when $\delta$ documents are returned while searching for $t$
  terms out of the $m$ total distinct terms that are indexed in the
  collection. This parameter affects the bandwidth usage of the system.

\item The computational effort at the server for constructing the
  proof. Let $N$ be the total size of the inverted lists of the query
  terms. The asymptotic running time at the server is
  $O(N\log^2N\log\log N)$~\cite{ptt-ovods-11}.
  Note that the overhead over the time needed to compute a plain set
  intersection, which is $O(N)$, is only a polylogarithmic
  multiplicative.  In practice, the critical computation at the server
  is the extended Euclidean algorithm, which is executed to construct
  the completeness witnesses.

\item The computational effort at the client to verify the proof.  The
  asymptotic running time at the client is $O(\delta + t\log m)$ for $t$
  query terms, $\delta$ documents in the query result and an inverted
  index of $m$ distinct terms.

\item The computational effort at the crawler to update the
  authenticated data structure when some documents are added to or
  deleted from the collection.  The asymptotic running time at the
  crawler consists of updating accumulation values and corresponding
  Merkle tree paths for $t'$ unique terms that appear in $n'$ updated
  documents, and, hence, it is $O(t'n' + t'\log m)$.

\end{enumerate}

\subsection{Prototype Implementation}

Our prototype is built in C++ and is split between three parties:
authenticated crawler, search engine and client.  The interaction
between the three proceeds as follows.

The crawler picks a secret key $s$ and processes a collection of
documents.  After creating the inverted index, the crawler computes an
accumulation value for each inverted list.  We use cryptographic pairing
from~\cite{nns-nssrcp-10} for all bilinear pairing computations which are
available in DCLXVI
library.\footnote{\url{http://www.cryptojedi.org/crypto/}}
This library implements an optimal ate pairing on a Barreto-Naehrig
curve over a prime field $\mathbb{F}_p$ of size 256 bits, which provides
128-bit security level.  Hence, the accumulation value of the documents
containing a given term is represented as a point on a curve.  Once all
accumulation values are computed, the crawler builds a Merkle tree where
each leaf corresponds to a term and its accumulation value.  We use
SHA256 from the OpenSSL library\footnote{\url{http://www.openssl.org/}}
to compute a hash value at each node of the Merkle tree.  The crawler
also computes values $g$, $g^s$, $\ldots$, $g^{s^n}$.
 
After the authenticated data structure is built, the crawler outsources
the inverted index, authentication layer and precomputed values $g$,
$g^s$, $\ldots$,$g^{s^n}$ to the server.

Clients query the search engine via the RPC interface provided by the
Apache Thrift software framework\ifFull ~\cite{lib-thrift}\fi.  For each
query, the server computes the proof consisting of four parts.  \ifFull
To efficiently compute the proof, the server makes use of the following
algorithms
\begin{itemize}
\item The coefficients $b_\delta, b_{\delta-1}, \ldots, b_{0}$ and the
  coefficients for the subset witnesses are computed using FFT.
\item The Extended Euclidean algorithm is used to compute the
  coefficients $q_1(s),$ $q_2(s), \ldots, q_{t}(s)$ for the completeness
  witnesses.
\end{itemize}
\fi 
We use the NTL\footnote{A Library for doing Number Theory, V5.5.2.}
and LiDIA\footnote{A library for Computational Number Theory, V2.3.}
libraries for efficient arithmetic operations, FFT and the Euclidean
algorithm on elements in $\mathbb{Z}^{*}_p$, which represent document
identifiers.  Bilinear pairing, power and multiplication operations on
the group elements are performed with the methods provided in the
bilinear map library.

We performed the following optimizations on the server.  The computation
of subset and completeness witnesses is independent of each other and
hence is executed in parallel.  We also noticed that the most expensive
part of the server's computation is the power operation for group
elements when computing subset and completeness witnesses.  Since the
order of these computations is independent from each other we run the
computation of $g^{P_j(s)}$ and $g^{q_j(s)}$ in parallel.

The client's verification algorithm consists of verifying the
accumulation values using the Merkle tree and running the bilinear
pairing procedure over the proof elements.

\ifShort
\begin{figure*}[t]
\centering
\begin{tabular}{ccc}
\includegraphics[width=55mm,angle=270]{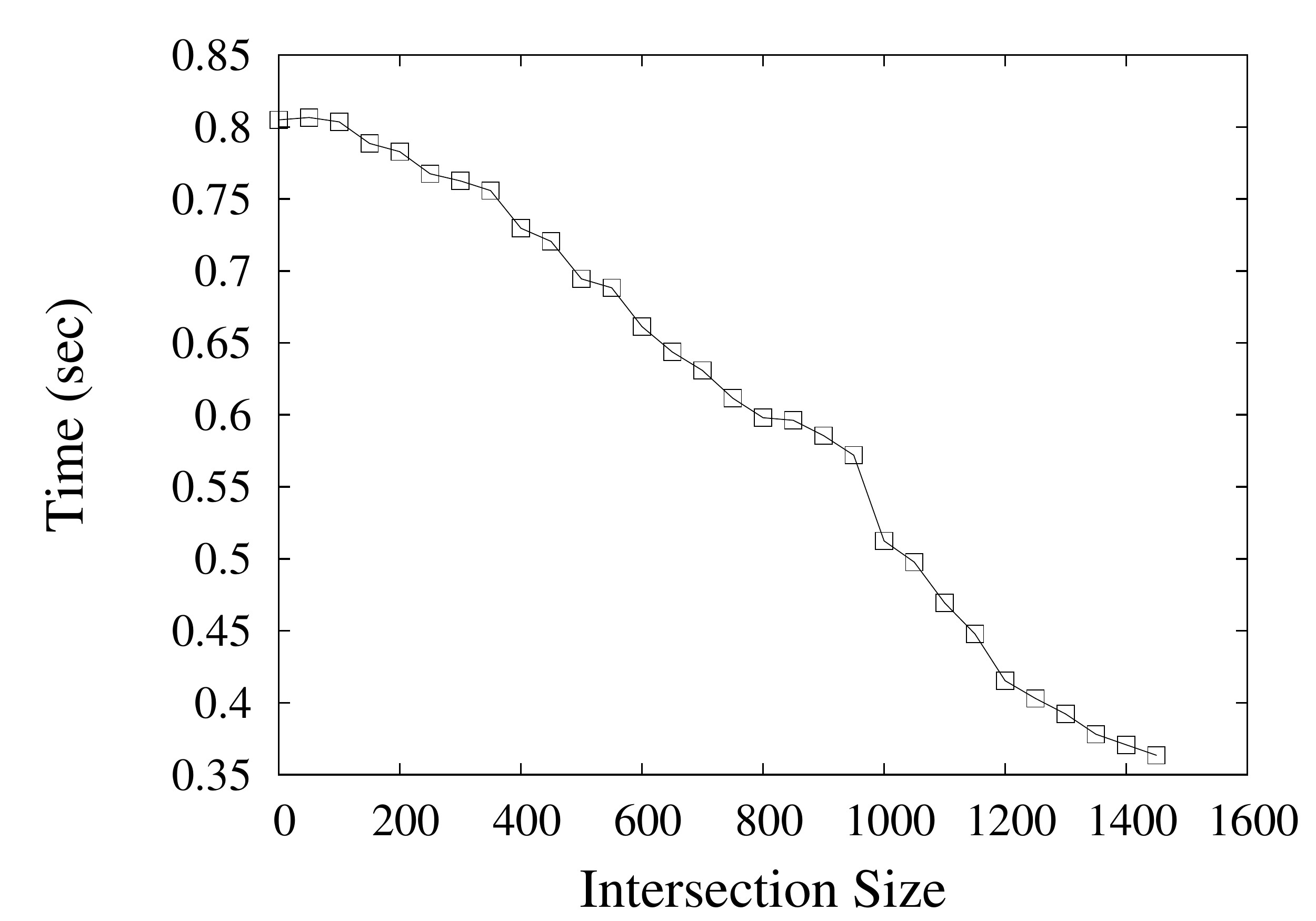}
\label{fig:synth_intsize_vs_time}
& \hspace{3mm} &
\includegraphics[width=55mm,angle=270]{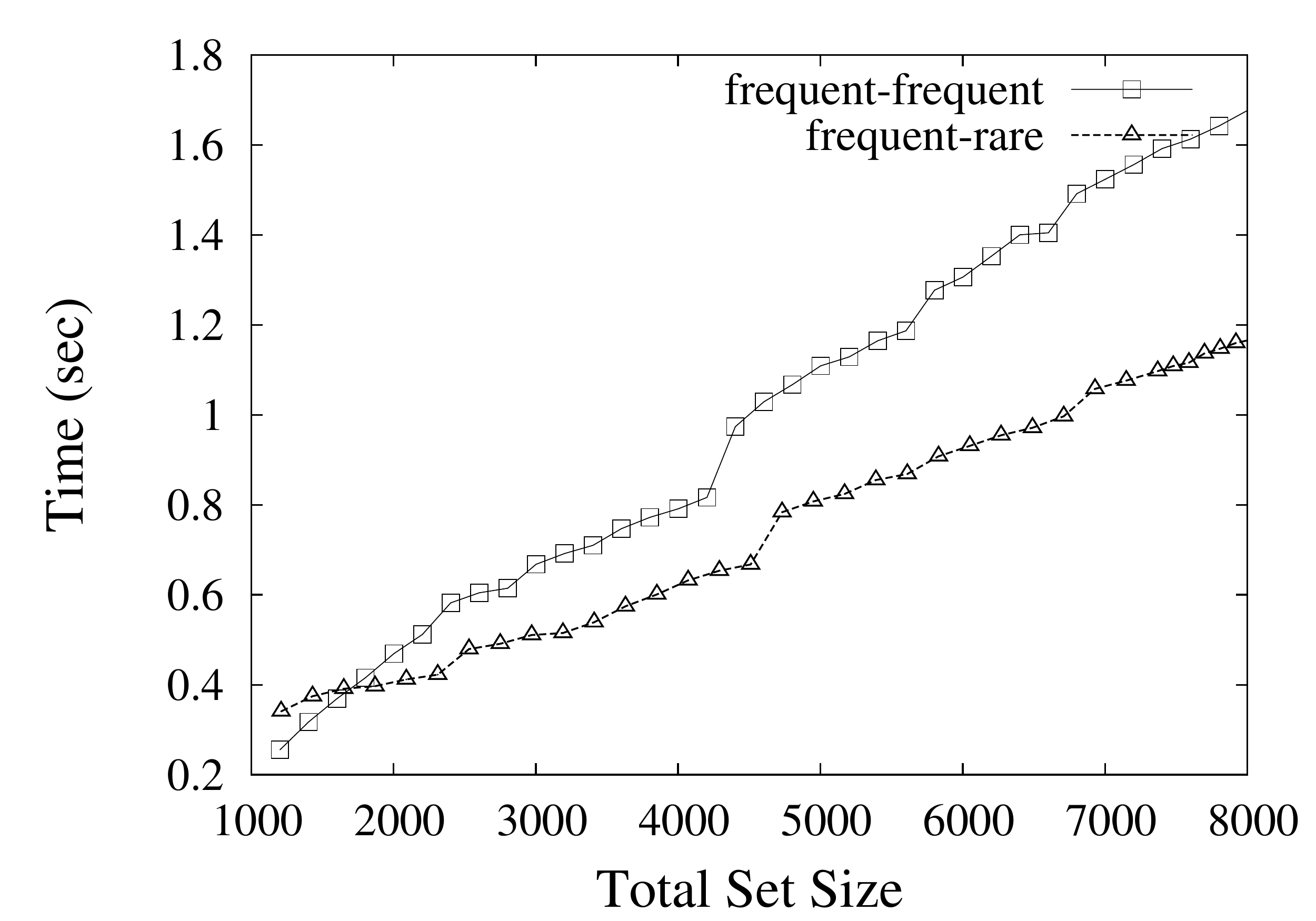}
\label{fig:synth_setsize_vs_time}
\end{tabular}
\caption{Computational effort at the server on synthetic data. (Left)
  Proof computation time for a 2-term query as a function of the
  intersection size (number of returned results) when each query term
  appears in 2,000 documents. (Right) Proof computation time as a function
  of the total set size (number of documents in queried inverted lists)
  for frequent-frequent and frequent-rare 2-term queries each returning
  100 documents.}
\end{figure*}
\fi

\subsection{Experimental Results}

We have conducted computational experiments on a 8-core Xeon 2.93
processor with 8Gb RAM running 64-bit Debian Linux. In the discussion of
the results, we use the following terminology.  
\ifFull
\begin{itemize}
\item \emph{Total set size}: sum of the lengths of the inverted lists of
  the query terms. This value corresponds to variable $N$ used in the
  asymptotic running times given in Section~\ref{sec:asymp_comp}.
\item \emph{Intersection size}: number of documents returned as the
  query result. This value corresponds to variable $\delta$ used in the
  asymptotic running times given in Section~\ref{sec:asymp_comp}.
\end{itemize}
\fi

\subsubsection{Synthetic Data}
\label{sec:results_synth_data}
We have created a synthetic data set where we can control the frequency
of the terms as well as the size of the intersection for each query.
The synthetic data set contains 777,348 documents and 320 terms.  Our
first experiment identifies how the size of the intersection, $\delta$,
influences the time it takes for the server to compute the proof given
that the size of the inverted lists stays the same. We report the
results in Figure~\ref{fig:synth_intsize_vs_time} where each point
corresponds to a query consisting of two terms. Each term used in a
query appears in 2,000 documents.  As the intersection size grows, the
size of the polynomial represented by the subset witness $P_j(s)$ in
Section~\ref{sec:solution_untrusted_server} decreases.  Hence the time
it takes the server to compute the proof decreases as well.

\ifFull
\begin{figure*}[t]
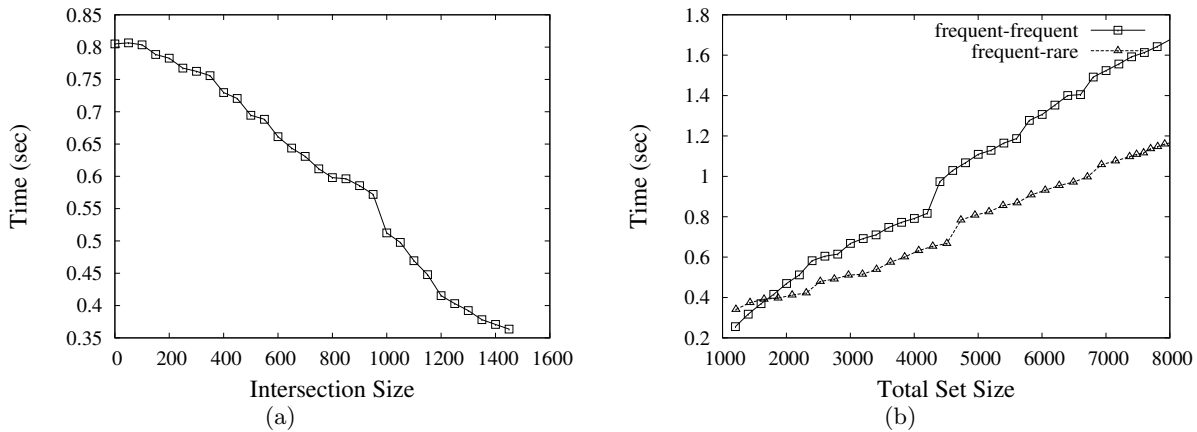

\subfigure[] {
\includegraphics[width=55mm,angle=270]{synthetic_intsize_vs_time}
\label{fig:synth_intsize_vs_time}
}
\subfigure[]{
\includegraphics[width=55mm,angle=270]{synthetic_setsize_vs_time}
\label{fig:synth_setsize_vs_time}
}
\caption{Computational effort at the server on a synthetic data set.
  (a)~Time to compute the proof for a query with two terms as a
  function of the intersection size, i.e., the number of returned
  query results. The length of the inverted list for each term is
  fixed at 2,000 documents.  (b)~Time to compute the proof for two
  types of queries: two frequent terms and frequent-rare term pair.
  The intersection size for each
  query, i.e., the number of returned documents, is~100.}
\end{figure*}
\fi

We now measure how the size of the inverted lists affects server's time
when the size of the resulting intersection is fixed to $\delta=100$
documents.  Figure~\ref{fig:synth_setsize_vs_time} shows results for
queries of two types.  \ifFull
\begin{itemize}
\item The first type consists of queries where both terms appear in the
  collection with the same frequency.
\item The second type of queries contain a frequent and a rare term.
\end{itemize}
\fi 
In each query we define a term as rare if its frequency is ten times
less than the frequency of the other term.  As the number of terms for
subset witnesses grows the time to compute these witnesses grows as
well. The dependency is linear, as expected (see
Section~\ref{sec:asymp_comp}). We also note that the computation is more
expensive when both terms in the query have the same frequency in the
collection.

\subsubsection{WSJ Corpus}
We have also tested our solution on a real data set that consists of
173,252 articles published in the Wall Street Journal from March 1987 to
March 1992. After preprocessing, this collection has 135,760 unique
terms.  We have removed common words, e.g., articles and prepositions,
as well as words that appear only in one document.
\ifFull 
The distribution of lengths of the inverted lists is shown in
Figure~\ref{fig:inv_lists} where 56\% of the terms appear in at most 10
documents.
\begin{figure}
\subfigure[]{
\includegraphics[width=55mm,angle=270]{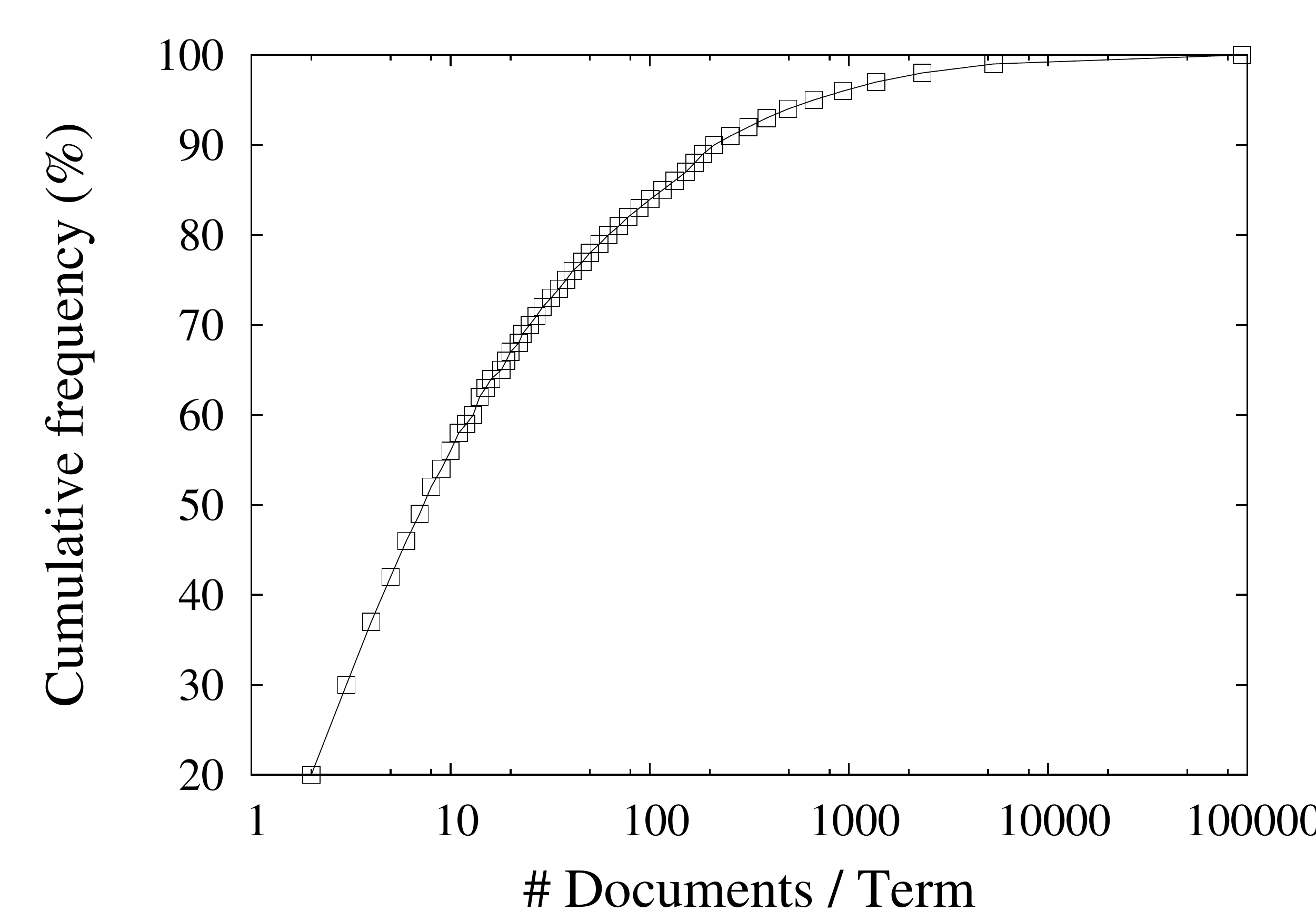}
\label{fig:inv_lists}}
\subfigure[]{
\includegraphics[width=55mm,angle=270]{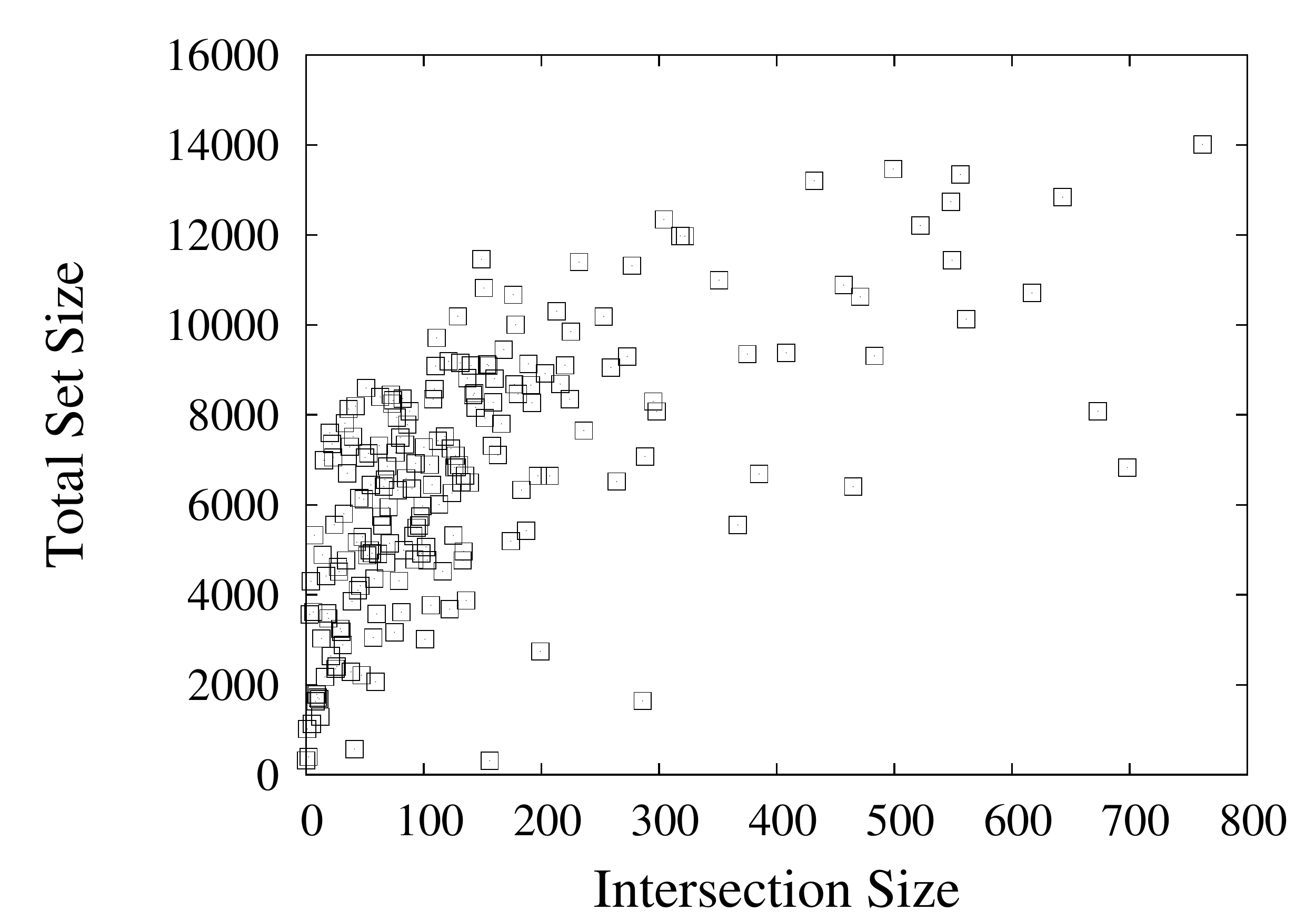}
\label{fig:intsize_vs_setsize}}
\caption{(a) Distribution of the length of the inverted lists for the
  WSJ corpus.  (b) Query set for WSJ corpus. Each point in the graph
  corresponds to a query and its intersection size in the corpus,
  i.e. the number of documents that have all terms from the query.}
\end{figure}
\fi

Our query set consists of 200 random conjunctive queries with two terms.
We picked queries that yield varying result sizes, from empty to 762
documents.  Since each term in a query corresponds to an inverted list
of the documents it appears in, we also picked rare as well as frequent
terms.  Here, we considered a term frequent if it appeared in at least
8,000 documents.  The total set size and corresponding intersection size
for each query is shown in Figure~\ref{fig:intsize_vs_setsize}.

\begin{figure*}[t!]
\ifShort
\centering
\begin{tabular}{ccc}
\includegraphics[width=55mm,angle=270]{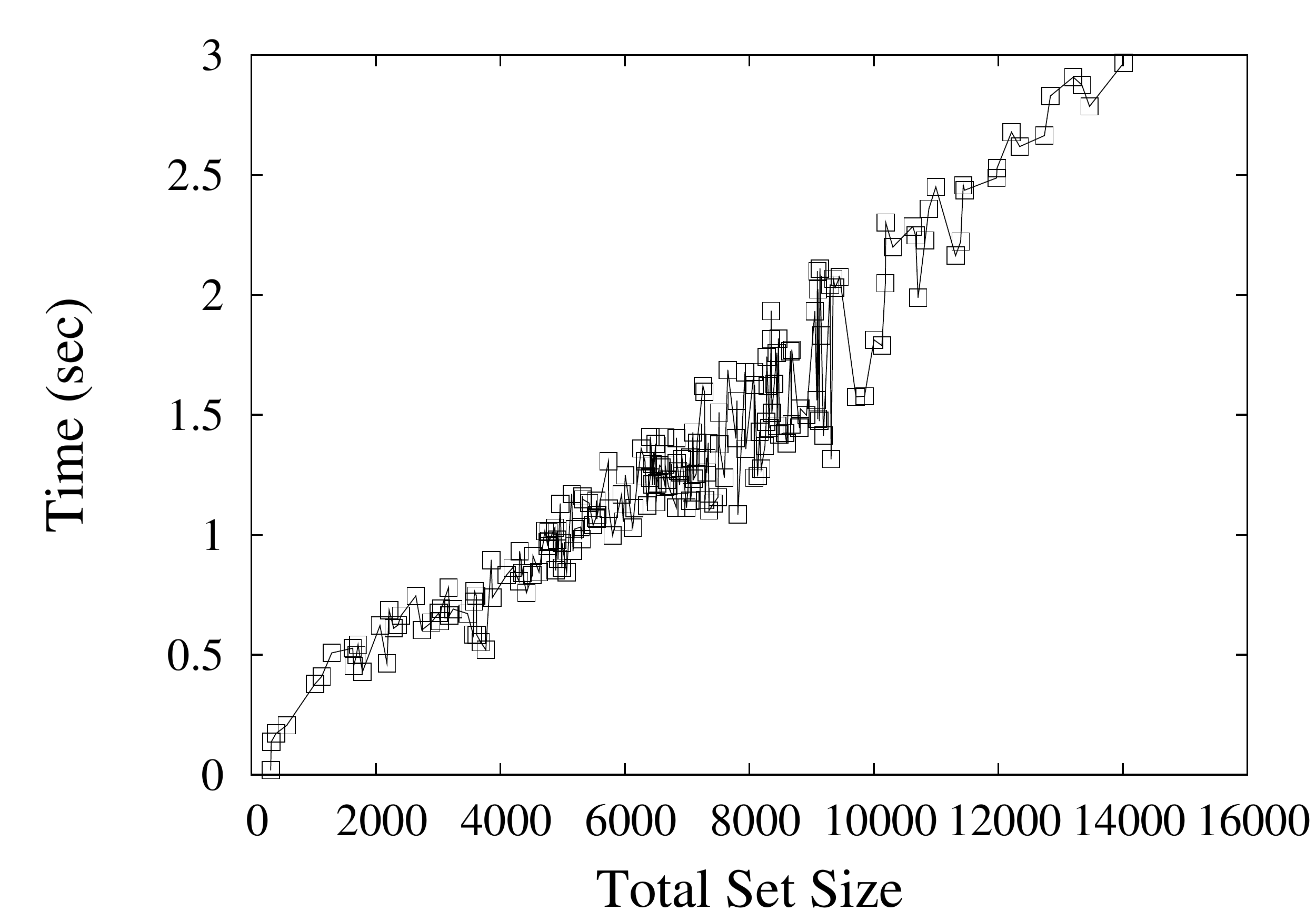}
\label{fig:time_vs_setsize}
& \hspace{3mm} &
\includegraphics[width=55mm,angle=270]{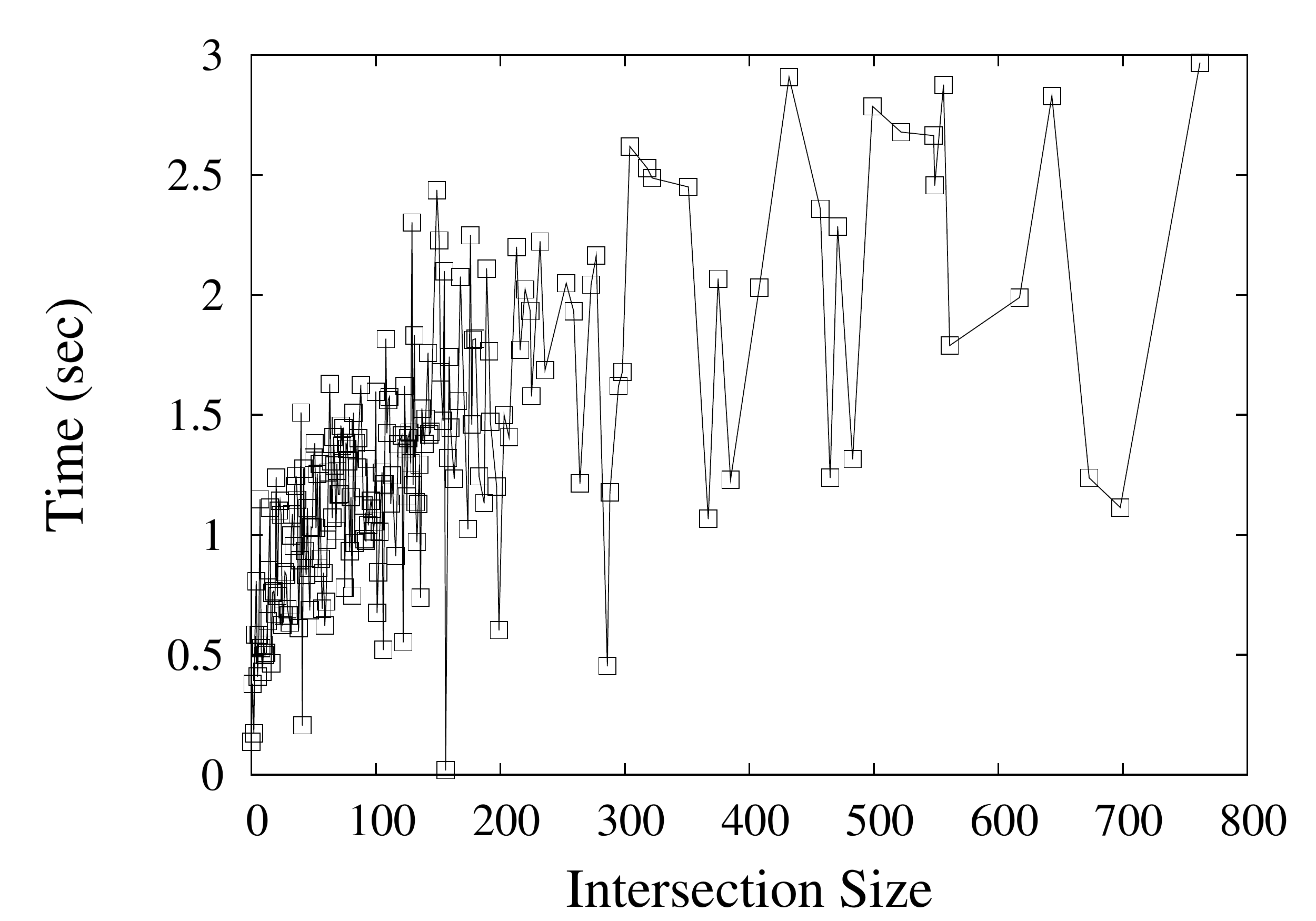}
\label{fig:time_vs_intsize}
\end{tabular}
\caption{Computational effort at the server for queries with two terms
  on the WSJ corpus. Time to compute the proof as a function of the
  total set size (left) and the intersection size (right).}
\label{fig:time_vs_size}
\else 
\subfigure[]{
\includegraphics[width=55mm,angle=270]{wsj_setsize_vs_time}
\label{fig:time_vs_setsize}
}
\subfigure[]{
\includegraphics[width=55mm,angle=270]{wsj_intsize_vs_time}
\label{fig:time_vs_intsize}
}
\caption{Computational effort at the server for queries with two terms
  on the WSJ corpus. Time to compute the proof as a function of (a) the
  total set size and (b) the intersection size.  The total set size is
  the total length the inverted lists of the query terms.}
\fi
\end{figure*}

\smallskip\noindent{\bf Server time.} We first measure the time it takes
for the server to compute the proof.
In \ifShort Figure~\ref{fig:time_vs_size} (left) \else Figure~\ref{fig:time_vs_setsize} \fi
we show how the size of the inverted
lists of the terms in the query influences server's time. As expected,
the dependency is linear in the total set size. However, some of the
queries that have inverted lists of close length result in different
times.
\ifShort
\begin{figure}[t]
\centering
\includegraphics[width=53mm,angle=270]{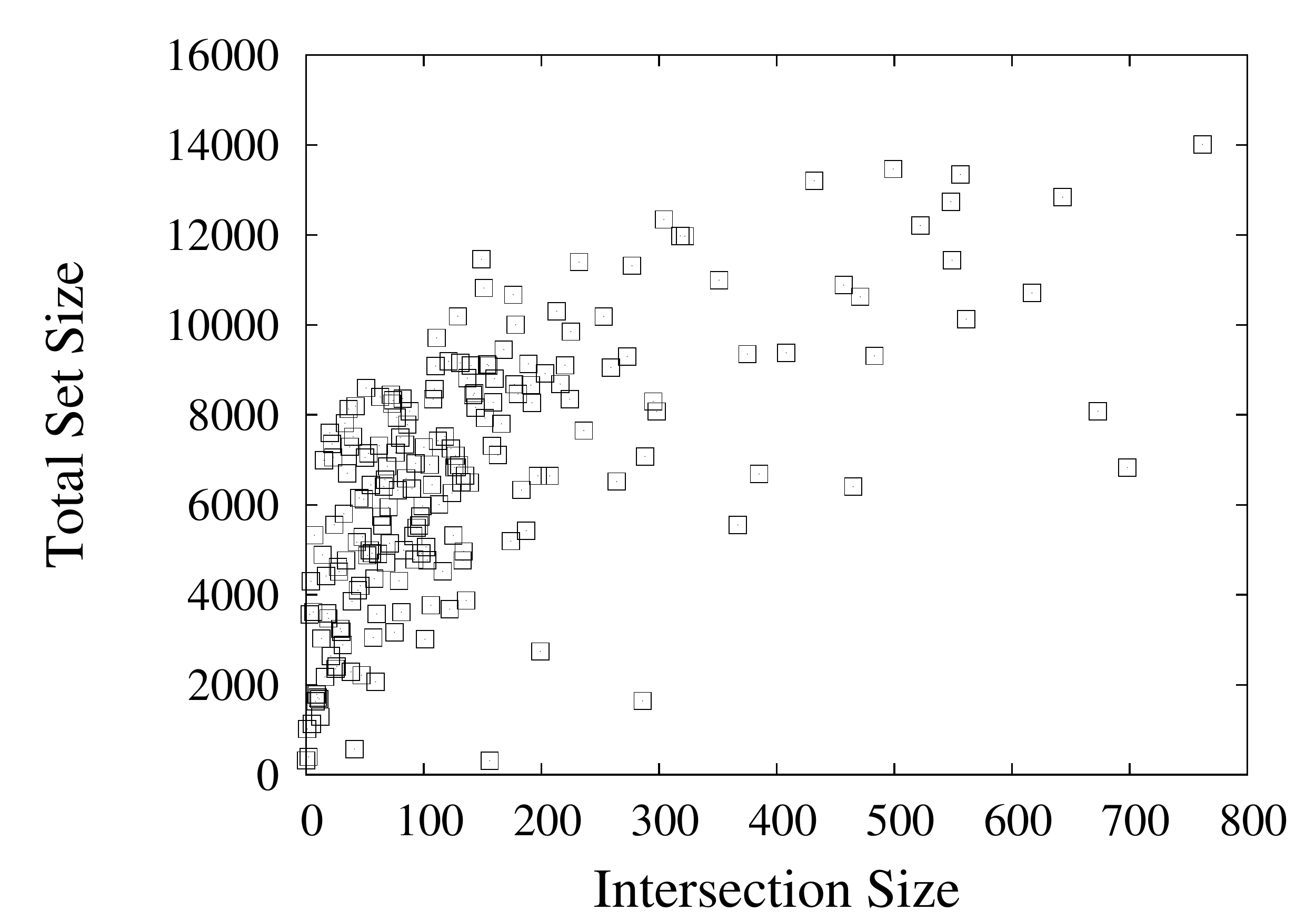}
\caption{Query set for WSJ corpus. Each point relates the number of
  returned documents to the number of documents in the queried inverted
  lists.}
\label{fig:intsize_vs_setsize}
\end{figure}
\fi
This happens because the intersection size varies between the queries,
as can be seen in Figure~\ref{fig:intsize_vs_setsize}.  Furthermore,
some of the queries contain different type of terms, e.g., consider a
query with one rare and one frequent word, and a query with two
semi-frequent words (see Section~\ref{sec:results_synth_data}).

In \ifShort Figure~\ref{fig:time_vs_size} (right) \else Figure~\ref{fig:time_vs_intsize} \fi we show how the intersection size
influences server's time. Note that the graph is almost identical to
Figure~\ref{fig:intsize_vs_setsize}, again showing that the time mostly
depends on the lengths of the inverted lists and not the intersection
size.

\smallskip\noindent{\bf Client time and proof size.} We now measure the
time it takes for the client to verify the proof.  Following the
complexity analysis of our solution, the computation on the client side
is very fast.  We split the computation time since verification of the
proof consists of verifying that intersection was performed on correct
inverted lists (Merkle tree) and that intersection itself is computed
correctly (bilinear pairing on accumulation values).  In
Figure~\ref{fig:intsize_vs_vtime}, we plot the time it takes to verify
both versus the intersection size: It depends only on the intersection
size and not on the total set size (the lengths of the inverted lists of
the query terms).
Finally, the size of the proof sent to the client is proportional to the
intersection size as can be seen in Figure~\ref{fig:intsize_vs_vsize}.

\begin{figure}[t!]
\centering
\ifFull
\includegraphics[width=55mm,angle=270]{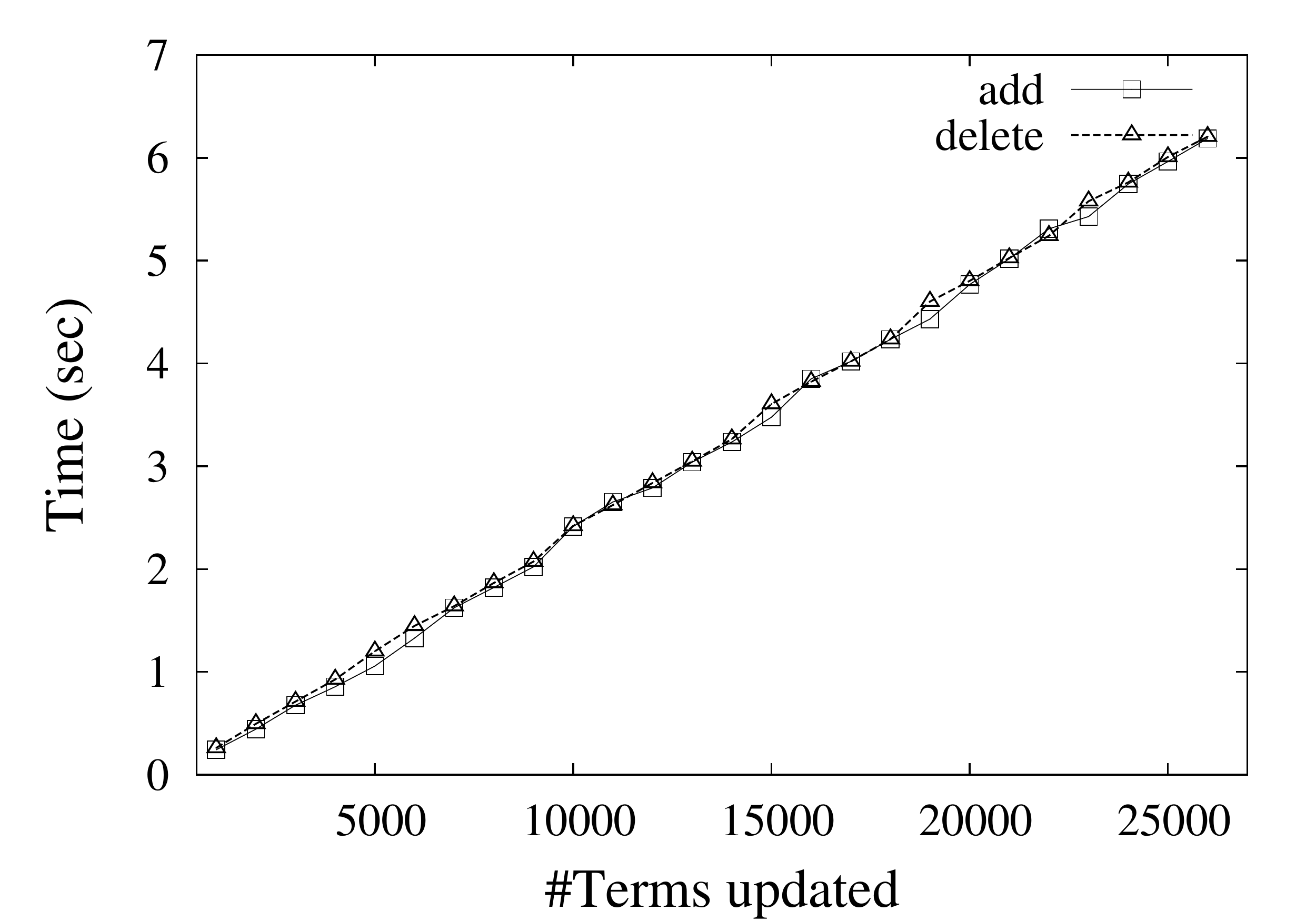}
\caption{Time it takes for the crawler to update the authenticated data
  when documents are added or deleted from the original collection.}
\label{fig:updates}
\fi
\end{figure}

\ifShort
\begin{figure}[t!]
\centering
\includegraphics[width=53mm,angle=270]{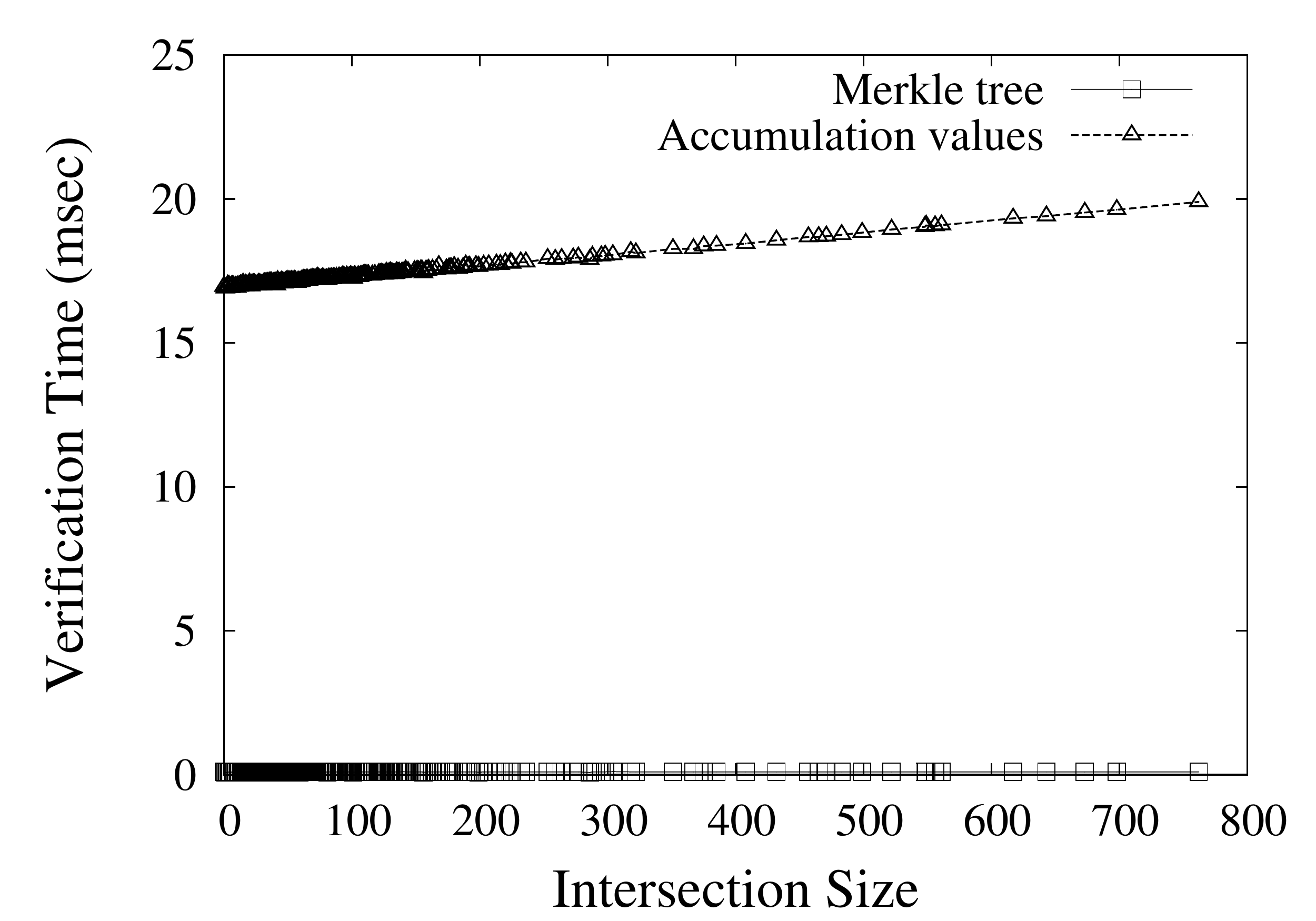}
\caption{Verification time at the client as a function of the
  intersection size split into its two components.}
\label{fig:intsize_vs_vtime}
\end{figure}
\begin{figure}[t!]
\centering
\includegraphics[width=53mm,angle=270]{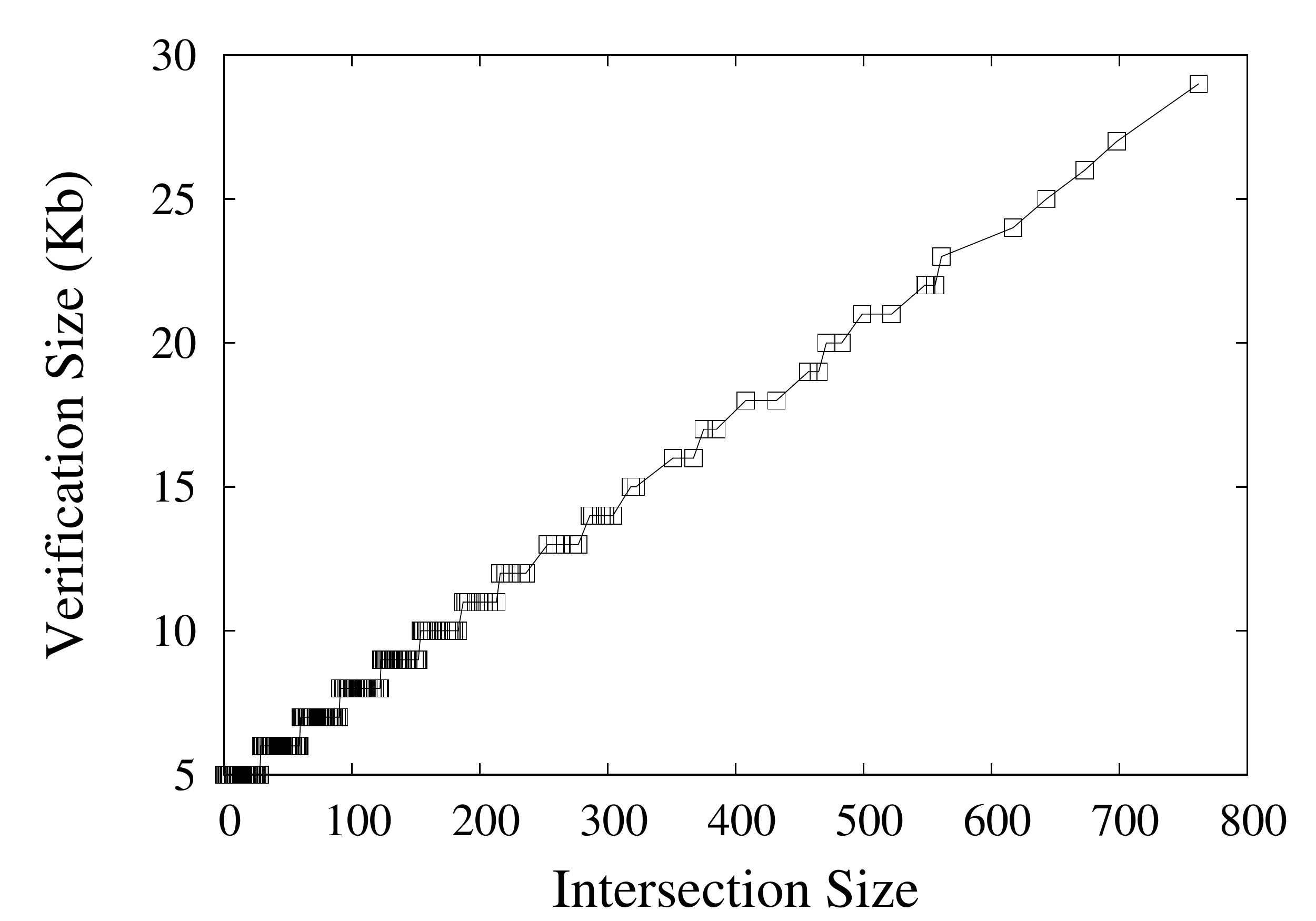}
\caption{Proof size as function of intersection size.}
\label{fig:intsize_vs_vsize}
\end{figure}
\else
\begin{figure}[t!]
\begin{minipage}[b]{0.5\linewidth}
\includegraphics[width=55mm,angle=270]{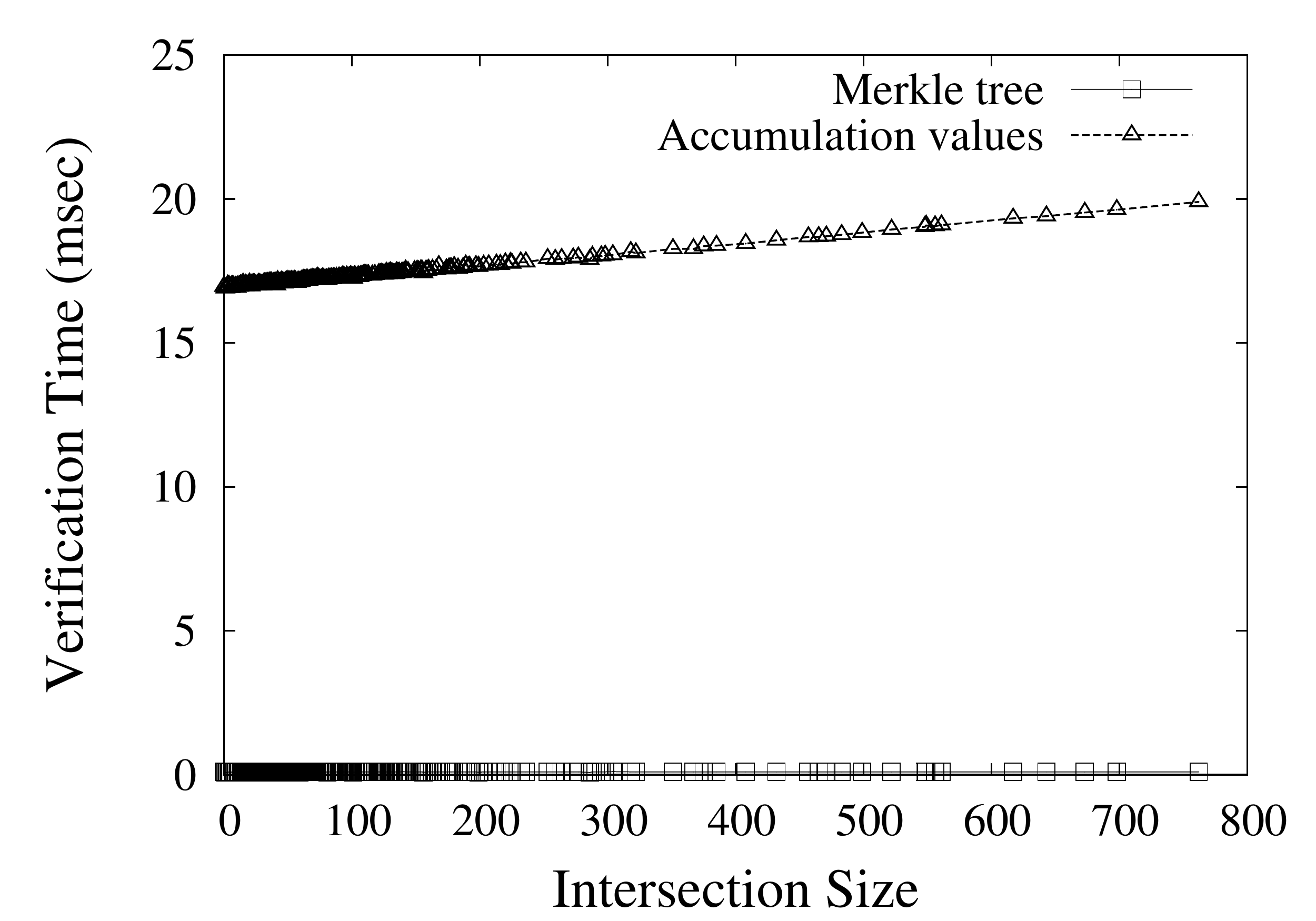}
\caption{Time at the client to verify the proof as a function of the
  intersection size. The time is split between verifying the leaves of
  Merkle tree and integrity of intersection using accumulation values.}
\label{fig:intsize_vs_vtime}
\vspace{-14mm}
\end{minipage}
\hspace{0.2cm}
\begin{minipage}[b]{0.5\linewidth}
\includegraphics[width=55mm,angle=270]{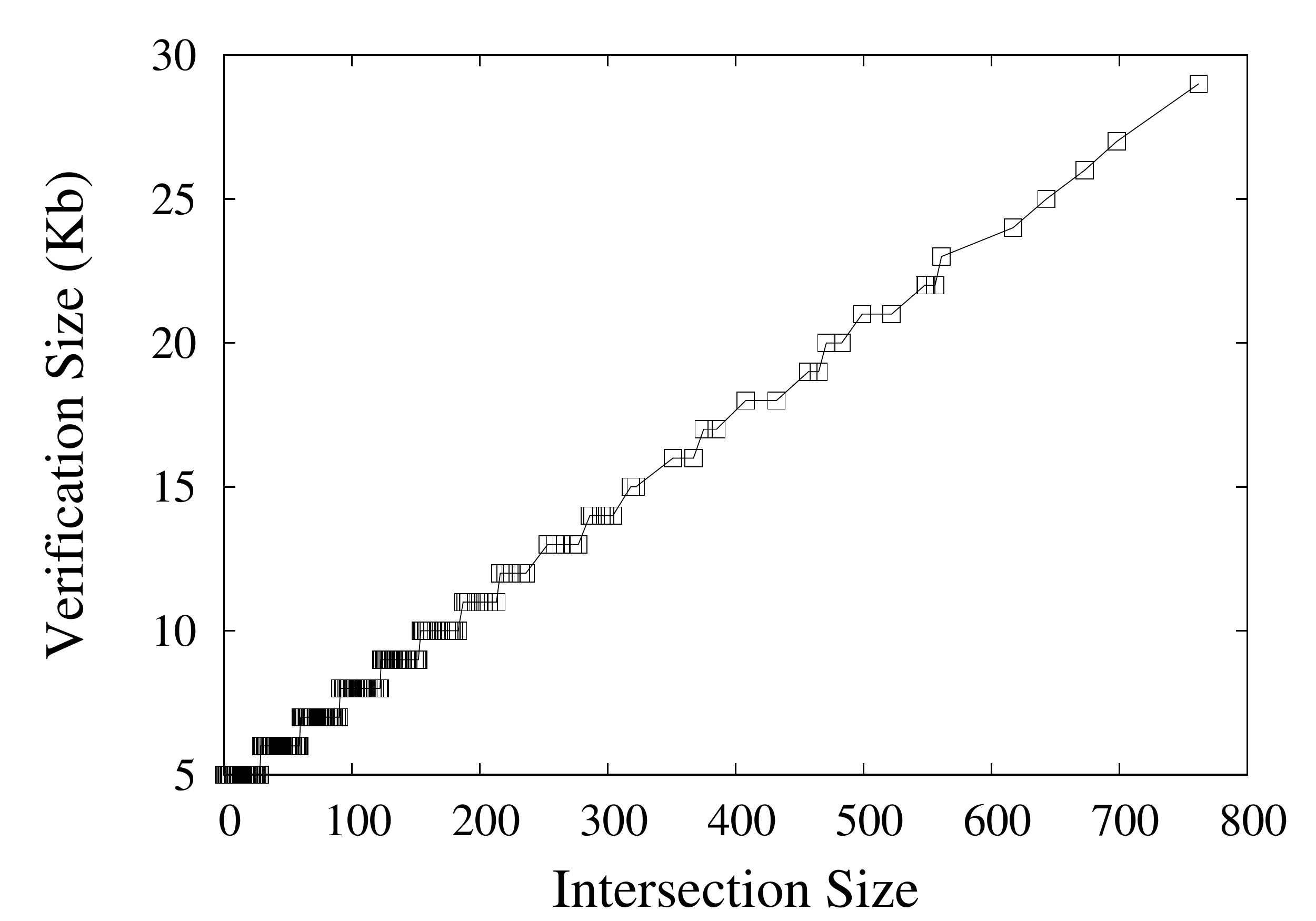}
\caption{Size of the proof as a function of the intersection size.}
\label{fig:intsize_vs_vsize}
\end{minipage}
\vspace{5mm}
\end{figure}
\fi
\smallskip\noindent{\bf Updates to the corpus.}
The simulation supports addition and deletion of new documents and
updates corresponding authenticated data structures.  We pick a set of
1500 documents from the original collection which covers over 14\% of
the collection vocabulary.  In Figure~\ref{fig:updates} we measure the
time it takes for the crawler to update accumulation values in the
authenticated data structure.  As expected the time to do the update is
linear in the number of unique terms that appear in the updated document
set.  Deletions and additions take almost the same time since the update
is dominated by a single exponentiation of the accumulation value of
each affected term.  Updates to Merkle tree take milliseconds.
%

\subsubsection{Comparison with Previous Work}

The closest work to ours is the method developed by Pang and
Mouratidis~\cite{pm-aqrtse-08}.  However, their method solves a
different problem.
Using our method, the user can verify whether the result of the query
contains all the documents from the collection that satisfy the query
and no malicious documents have been added. The method
of~\cite{pm-aqrtse-08} proves that the returned result consists of
top-ranked documents for the given query. However, it does not assure
the completeness of the query result with respect to the query terms.

The authors of~\cite{pm-aqrtse-08} show that their best technique
achieves below 60 msec verification time,\footnote{Dual Intel Xeon
  3GHz CPU with 512MB RAM machine.}  and less than 50 Kbytes in proof
size for a result consisting of 80 documents.  Using our method the
verification time for a result of 80 documents takes under 17.5 msec
(Figure~\ref{fig:intsize_vs_vtime}) and the corresponding verification
object is of size under 7 Kbytes (Figure~\ref{fig:intsize_vs_vsize}).
The computation effort by the server reported in~\cite{pm-aqrtse-08} is
lower than it is for our method, one second versus two seconds.

We also note that updates for the solution in~\cite{pm-aqrtse-08}
require changes to the whole construction, while updates to our
authenticated data structures are linear in the number of unique terms
that appear in new documents.

\subsubsection{Improvements and Extensions}
From our experimental results, we observe that the most expensive part
of our solution is the computation of subset and completeness witnesses
at the server.  This is evident when a query involves frequent terms
with long inverted lists, where each term requires a call to a
multiplication and power operation of group elements in $\mathbb{G}$.
However, these operations are independent of each other and can be
executed in parallel.  Our implementation already runs several
multiplication operations in parallel. However, the number of parallel
operations is limited on our 8-core processor.

In a practical deployment of our model, the server is in the cloud and
has much more computational power than the client, e.g., the server is a
search engine and the client is a portable device.  Hence, with a more
powerful server, we can achieve faster proof computation for frequent
terms.

Our implementation could use a parallel implementation of the Extended
Euclidean Algorithm, however, the NTL library
is not thread-safe and therefore we could not perform this optimization
for our current prototype.



\section{Conclusion}
\label{sec:conclusion}
We study the problem of verifying the results of a keyword-search query
returned by a search engine. We introduce the concept of an
authenticated web-crawler which enables clients to verify that the query
results returned by the search engine contain all and only the web pages
satisfying the query.  Our prototype implementation has low
communication overhead and provides fast verification of the results.


Our method verifies the correctness and completeness of the results but
does not check the ranking of the results. An interesting extension to
our method would be to efficiently verify also the ranking, i.e., return
to the client $r$ pages and a proof that the returned set consists of
the top-$r$ results.
\ifShort
\balance
\fi

\ifFull
\subsection*{Acknowledgments}
\fi

This research was supported in part by the National Science Foundation
under grants CNS-1012060, CNS-1012798, CNS-1012910, OCI-0724806 and
CNS-0808617, by a \mbox{NetApp} Faculty Fellowship, and by Intel through
the ISTC for Secure Computing.
We thank Isabel Cruz, Kristin Lauter, James Lentini, and Michael Naehrig
for useful discussions.


\bibliographystyle{abbrv}

\ifFull
\ifArxiv
\bibliography{main_arxiv,libraries}
\else
\bibliography{main,libraries,geom,crypto,goodrich,dsa,adaptive,amicExtra} 
\fi
\fi

\end{document}